\documentclass[12pt, a4paper]{article} 
\usepackage[utf8]{inputenc}
\usepackage{dcolumn,lscape}
\usepackage{amsmath,longtable,multicol,dcolumn,tabularx,graphicx,amssymb}
\usepackage{exscale,amsthm,multirow,rotating}
\usepackage{natbib}
\usepackage{bbm}
\usepackage{tikz,dsfont,float}
\usepackage{subcaption}
\definecolor{blue}{RGB}{0,0,0}
\usetikzlibrary{calc}
\usetikzlibrary{positioning}

\usepackage{graphicx}
\usepackage{amsmath,amssymb,dsfont, mathtools}
\usepackage{natbib}
\usepackage{makecell}
\usepackage{multirow}
\usepackage{exscale,longtable,multicol,dcolumn,tabularx,bm}
\usepackage{adjustbox}
\usepackage{booktabs} 
\usepackage{mathptmx}      
\usepackage{mathrsfs}
\DeclareMathOperator{\EX}{\mathbb{E}}
\newcommand{\indep}{\perp \!\!\! \perp}
\usepackage{float}

\usepackage{xcolor, ulem}
\begin{document}

\def\spacingset#1{\renewcommand{\baselinestretch}%
{#1}\small\normalsize} \spacingset{1}


  \title{\bf Online network monitoring}
  \author{Anna Malinovskaya\\
  	\small{Leibniz University Hannover, Germany}\\
  	Philipp Otto\\
  	\small{Leibniz University Hannover, Germany}}
  \maketitle
\begin{abstract}
The application of network analysis has found great success in a wide variety of disciplines; however, the popularity of these approaches has revealed the difficulty in handling networks whose complexity scales rapidly. One of the main interests in network analysis is the online detection of anomalous behaviour. To overcome the curse of dimensionality, we introduce a network surveillance method bringing together network modelling and statistical process control. Our approach is to apply multivariate control charts based on exponential smoothing and cumulative sums in order to monitor networks determined by temporal exponential random graph models (TERGM). This allows us to account for temporal dependence, while simultaneously reducing the number of parameters to be monitored. The performance of the proposed charts is evaluated by calculating the average run length for both simulated and real data. To prove the appropriateness of the TERGM to describe network data, some measures of goodness of fit are inspected. We demonstrate the effectiveness of the proposed approach by an empirical application, monitoring daily flights in the United States to detect anomalous patterns.
\end{abstract}

\noindent%
\vfill

\spacingset{1.45} 

\section{Introduction}
\label{intro}
The digital information revolution offers a rich opportunity for scientific progress; however, the amount and variety of data available requires new analysis techniques for data mining, interpretation and application of results to deal with the growing complexity. As a consequence, these requirements have influenced the development of networks, bringing their analysis beyond the traditional sociological scope into many other disciplines, as varied as are physics, biology and statistics (cf. \citealt{amaral2000classes,simpson_bowman_laurienti_2013,chen_haerdle_okhrin_2019}).

One of the main interests in network study is the detection of anomalous behaviour. There are two types of network monitoring, differing in the treatment of nodes and links: fixed and random network surveillance (cf. \citealt{leitch_alexander_sengupta_2019}). {\color{blue} In this work,} we concentrate on the modelling and monitoring of networks with randomly generated edges across time, describing a surveillance method of the second type. When talking about anomalies in temporal networks, the major interest is to find the point of time when a significant change happened and, if appropriate, to identify the vertices, edges or graph subsets which considerably contributed to the change (cf. \citealt{akoglu_tong_koutra_2014}). Further differentiating depends on at least two factors: characteristics of the network data and available time granularity. Hence, given a particular network to monitor it is worth first defining what is classified as ``anomalous''.

To analyse the network data effectively and plausibly, it is important to account for its complex structure and the possibly high computational costs. Our approach to mitigate these issues and simultaneously reflect the stochastic and dynamic nature of networks is to model them applying a temporal random graph model. We consider a general class of Exponential Random Graph Models (ERGM) (cf. \citealt{frank_strauss_1986,robins2007introduction,schweinberger2017exponential}), which was originally designed for modelling cross-sectional networks. This class includes many prominent random network configurations such as dyadic independence models and Markov random graphs, enabling the ERGM to be generally applicable to many types of complex networks. {\color{blue} Furthermore, if a network has many attributes, i.e., variables that provide additional information about the graph’s elements, it can be both computationally and theoretically challenging to recognise meaningful patterns. In this case, it is beneficial to apply ERGM as it compactly summarises the knowledge about the network in several terms.} \cite{hanneke2010discrete} developed a dynamic extension based on ERGM {\color{blue} which is known as} the Temporal Exponential Random Graph Model (TERGM). These models contain the overall functionality of the ERGM, while additionally enabling time-dependent covariates.  Thus, our monitoring procedure for this class of models allows for many applications in different disciplines which are interested in analysing networks of medium sizes, such as sociology, political science, engineering, economics and psychology (cf. \citealt{carrington2005models,ward2011network,das2013topological,jackson2015past,fonseca2018network}).

In the field of change detection, according to \cite{basseville1993detection} there are three classes of problems: online detection of a change, off-line hypotheses testing and off-line estimation of the change time. Our method refers to the first class, meaning that the change point should be detected as soon as possible after it occurred. In this case, real-time monitoring of complex structures becomes necessary: for instance, if the network is observed every minute, the monitoring procedure should be faster than one minute. To perform online surveillance for real-time detection, the efficient way is to use tools from the field of Statistical Process Control (SPC). SPC corresponds to an ensemble of analytical tools originally developed for industrial purposes, which is applied for the achievement of process stability and variability reduction (e.g., \citealt{montgomery2012statistical}).

The leading SPC tool for analysis is a control chart, which exists in various forms in terms of the number of variables, data type and statistics being of interest. For example, the monitoring of network topology statistics applying the Cumulative Sum (CUSUM) chart and illustrating its effectiveness on the analysis of military networks was presented by \cite{mcculloh2011detecting}. \cite{wilson2019modeling} used the dynamic Degree-Corrected Stochastic Block Model (DCSBM) to generate the networks and then performed surveillance over the Maximum Likelihood (ML) estimates using the Shewhart and Exponentially Weighted Moving Average (EWMA) charts. One of the possibilities to bring together the ERGM in form of a Markov Graph and EWMA and Hotelling's $T^2$ charts was proposed by \cite{sadinejad2020monitoring}. \cite{farahani2017statistical} evaluate the combination of Multivariate EWMA (MEWMA) and Multivariate CUSUM (MCUSUM) applied with the Poisson regression model for monitoring social networks. \cite{hosseini2018performance} apply EWMA and CUSUM to degree measures for detecting outbreaks on a weighted undirected network. The distribution-free MCUSUM is introduced by \cite{liu2019analyzing} to analyse longitudinal networks. \cite{salmasnia2019change} present a comparative study of univariate and multivariate EWMA for social network monitoring. An overview of further studies is provided by \cite{noorossana2018overview}.

In this paper, we present an online monitoring procedure based on the SPC, which enables one to detect significant changes in the network structure in real time. The foundations of this approach together with the description of the selected network model and multivariate control charts are discussed in Section \ref{section:2}. Section \ref{section:3} outlines the simulation study and includes performance evaluation of the designed control charts. In Section \ref{section:4} we monitor daily flights in the United States and explain the detected anomalies. We conclude with a discussion of outcomes and present several directions for future research.

\section{Network Monitoring}\label{section:2}
Network monitoring is a form of an online surveillance procedure to detect deviations from a so-called in-control state, i.e., the state when no unaccountable variation of the process is present. This is done by sequential hypothesis testing over time, which has a strong connection to control charts. In other words, the purpose of control charting is to identify occurrences of unusual deviation of the observed process from a prespecified target (or in-control) process, distinguishing common from special causes of variation (cf. \citealt{johnson2007applied}).  To be precise, the aim is to test the null hypothesis
\begin{equation*}
H_{0, t}: \, \text{The network observed at time point $t$ is in its {\color{blue}target} state}
\end{equation*}
against the alternative
\begin{equation*}
\hspace*{1.75cm}H_{1, t}: \, \text{The network observed at time point $t$ deviates from its {\color{blue}target} state.}
\end{equation*}
In this work, we concentrate on the monitoring of networks, which are modelled by the TERGM that is briefly described below.

\subsection{Network Modelling}

The network (also interchangeably called ``graph'') is presented by its adjacency matrix \linebreak $\bm{Y} \coloneqq (Y_{ij})_{i,j = 1, \dots, N}$, where $N$ represents the total number of nodes. Two vertices (or nodes) $i, j$ are adjacent if they are connected by an edge (also called a tie or link). In this case, $Y_{ij} = 1$, otherwise, $Y_{ij} = 0$. In case of an undirected network, $\bm{Y}$ is symmetric. The connections of a node with itself are mostly not applicable to the majority of the networks, therefore, we assume that $Y_{ii} = 0$ for all $i = {1, \dots, N}.$
Formally, we define a network model as a collection $\{\mathbb{P}_{\theta}(\bm{Y}),\  \bm{Y} \in \mathscr{Y} : \bm{\theta} \in \Theta\}$, where $\mathscr{Y}$ denotes the ensemble of possible networks, $\mathbb{P}_{\theta}$ is a probability distribution on $\mathscr{Y}$ and $\bm{\theta}$ is a vector of parameters, ranging over possible values in {\color{blue}a subset of $p$-dimensional Euclidean space} $\Theta \subseteq {\rm I\!R}^p$ with $p \in {\rm I\!N}$ \citep{kolaczyk_2009}. {\color{blue}In case of a directed graph, where the edges have a direction assigned to them,}  this stochastic mechanism determines which of the $N(N-1)$ edges {\color{blue}are present}, i.e., it assigns probabilities
to each of the $2^{N(N-1)}$ graphs (cf. \citealp{cannings2003models}).

The ERGM functional representation is given by
\begin{equation}
\mathrm{P}_{\bm{\theta}} (\bm{Y})= \dfrac{\exp[\bm{\theta}'\bm{s}(\bm{Y})]}{c(\bm{\theta})},
\label{eq:1}
\end{equation}
where $\bm{Y}$ is the adjacency matrix of an observed graph with $\bm{s} \ : \ \mathscr{Y} \rightarrow {\rm I\!R}^p$ being a $p$-dimensional statistic describing the essential properties of a network based on $\bm{Y}$ (cf. \citealp{frank1991statistical, wasserman_pattison_1996}). There are several types of network terms, including dyadic dependent terms, for example, a statistic capturing transitivity, and dyadic independent terms, for instance, a term describing graph density \citep{morris2008specification}. {\color{blue} It is also possible to include nodal and edge attributes into the statistics. The exact variety is influenced by whether we have a directed or undirected network. Although the overall concept presented in this work is valid for both graph types, we explicitly consider directed graphs from now on. } 

The {\color{blue}model} parameters $\bm{\theta}$ can be defined as respective coefficients of $\bm{s}(\bm{Y})$ which are of considerable interest in understanding the structural properties of a network. They reflect, on the network-level, the tendency of a graph to exhibit certain sub-structures relative to what would be expected from a model by chance, or, on the tie-level, the probability to observe a specific edge, given the rest of the graph \citep{block2018change}. The last interpretation follows from the representation of the problem as a log-odds ratio. The normalising constant in the denominator ensures that the sum of probabilities is equal to one, meaning it includes all possible network configurations $c(\bm{\theta}) = \sum_{\bm{Y}\in \mathscr{Y}}[\exp\bm{\theta}'\bm{s}(\bm{Y})]$ in the ensemble $\mathscr{Y}$.

In dynamic network modelling, a random sequence of {$\bm{Y}_{t}$} for {$t = 1, 2, \dots$} with $\bm{Y}_{t} \in \mathscr{Y}$ defines a stochastic process for all $t$. It is possible that the {\color{blue}order} of a graph differs across the time stamps. {\color{blue}Unlike the relational event models, where the edges are modelled without duration (cf. \citealp{butts20084}), in this work, we contemplate edges with duration.} To conduct surveillance over $\bm{Y}_{t}$, we propose to consider only the dynamically estimated {\color{blue} characteristics} of a {\color{blue}graph} in order to reduce computational complexity and to allow for real-time monitoring. In most of the cases, the dynamic network models serve as an extension of well-known static models. Similarly, the discrete temporal expansion of the ERGM is known as TERGM (cf. \citealp{hanneke2010discrete}) and can be seen as further advancement of a family of network models proposed by \cite{robins2001random}.

The TERGM defines the probability of a network at the discrete time point $t$ both as a function of counted subgraphs in $t$ and by including the network terms based on the previous graph observations until the particular time point $t - v$, that is
\begin{equation}
\mathrm{P}_{\bm{\theta}}(\bm{Y}_t|\bm{Y}_{t-1}, \dots, \bm{Y}_{t-v}, \bm{\theta})= \dfrac{\exp[\bm{\theta}'\bm{s}(\bm{Y}_t, \bm{Y}_{t-1},\dots, \bm{Y}_{t-v})]}{c(\bm{\theta}, \bm{Y}_{t-1}, \dots, \bm{Y}_{t-v})},
\label{eq:2}
\end{equation}
where $v$ represents the maximum temporal lag, capturing the networks which are incorporated into the $\bm{\theta}$ estimation at $t$, hence, defining the complete temporal dependence of $\bm{Y}_t$ {\color{blue}that corresponds to the Markov structure of order $v \in {\rm I\!N}$ \citep{hanneke2010discrete}. In Sections \ref{section:3} and \ref{section:4}, we assume $v = 1$, leading to $(\bm{Y}_t \indep \{\bm{Y}_1, \dots, \bm{Y}_{t-2}\} | \bm{Y}_{t-1})$.}

{\color{blue}To model the joint probability of $z$ networks between the time stamps $v + 1$ and $v + z$, we define $\mathrm{P}_{\bm{\theta}}$ based on Equation (\ref{eq:2})
	\begin{equation}
	\mathrm{P}_{\bm{\theta}}(\bm{Y}_{v + 1}, \dots, \bm{Y}_{v + z}|\bm{Y}_{1}, \dots, \bm{Y}_{v}, \bm{\theta}) = \prod_{t = v + 1}^{v + z} \mathrm{P}_{\bm{\theta}}(\bm{Y}_{t}| \bm{Y}_{t-1}, \dots, \bm{Y}_{t - v}, \bm{\theta}).
	\label{eq:3}
	\end{equation}
	It is worth mentioning that Equation (\ref{eq:3}) is valid if $v$ is correctly chosen, so that the networks \linebreak $\bm{Y}_{v + 1}, \dots, \bm{Y}_{v + z}$ are conditionally independent \citep{leifeld2018temporal}.}

Regarding the network statistics in the TERGM, $\bm{s}(\cdot)$ includes ``memory terms'' such as dyadic stability or reciprocity \citep{leifeld2018temporal}. {\color{blue}To distinguish the processes leading to the dissolution and formation of links, \cite{krivitsky2014separable} presented Seperable TERGM (STERGM). To be precise, the STERGM is a subclass of the TERGM class, which can reproduce any transition process captured by $\bm{\theta} = (\theta^+, \theta^-)$ and $\bm{s} = (\bm{s}^+, \bm{s}^-)$.}

The creation of a meaningful configuration of sufficient network statistics replicates its ability to represent and reproduce the observed network close to reality.  Its dimension can differ over time, however, we assume that in each time stamp $t$ we have the same network statistics  $\bm{s}(\cdot)$. In general, the selection of terms extensively depends on the field and context, although the statistical modelling standards such as avoidance of linear dependencies among the terms should be also considered  \citep*{morris2008specification}. {\color{blue} It is also helpful to perform goodness of fit tests, which enable one to find a compromise between the model's complexity and its explanatory power.} An improper selection  can often lead to a degenerate model, i.e., when the algorithm does not converge consistently (cf. \citealp{handcock2003assessing}; \citealp{schweinberger2011instability}).  In this case, as well as fine-tuning the configuration of statistics, one can modify some settings which design the estimation procedure of the model parameter, for example, the run time, the sample size or the step length \citep{morris2008specification}. Another {\color{blue} possible improvement} would be to add some robust statistics such as Geometrically-Weighted Edgewise Shared Partnerships (GWESP) \citep{snijders_pattison_robins_handcock_2006}. However, the TERGM is less prone to degeneracy issues {\color{blue}compared to the ERGM} as ascertained by  \cite{hanneke2010discrete}  and  \cite{leifeld2015theoretical}. {\color{blue}Overall}, we assume that most of the network surveillance studies can reliably estimate beforehand the type of anomalies which are possible to occur. This assumption guides the choice of terms in the models throughout the work.

\subsection{{\color{blue}Monitoring Process}}
{\color{blue} Although the monitoring procedure can be constructed by supervising $\bm{Y}_t$ directly, this approach is prone to become computationally intricate as it depends on the order of a graph, leading to the curse of dimensionality. In the case of TERGM, we believe there are two reasonable choices of network monitoring, namely it can be performed either in terms of the (normalised) network statistics or the model parameters whose dimension remains independent from the network evolvement.}

{\color{blue}To obtain a time series of the corresponding estimates, we propose to apply a moving window approach with the window size $z$. More precisely, we take into account the past $z$ observations of the network $\{\bm{Y}_{t-z+1}, \dots, \bm{Y}_t\}$ to estimate the respective quantities at time point $t$.
	
	Let $\bm{\theta}$ be the true model parameters and $\hat{\bm{\theta}}_t$ their estimates at time point $t$ based on the last $z$ network states. Similarly, the expected value of the network statistics $\EX_{\bm{\theta}}(\bm{s}(\bm{Y}))$ can be estimated as 
	\begin{equation}
	\hat{\bm{s}}_t = \frac{1}{z}\sum_{n=0}^{z-1}\bm{s}(\bm{Y}_{t-n}).
	\label{eq4}
	\end{equation}
	Note that the temporal network terms at time point $t-n$ are calculated with respect to the temporal lag $v$. 
	
	Since the monitoring procedure is identical for $\hat{\bm{\theta}}_t$ and $\hat{\bm{s}}_t$, we introduce a new notation $\hat{\bm{c}}_t$ for the estimates of the network characteristics. Consequently, we refer to $\bm{c}$ meaning either $\bm{\theta}$ or $\EX_{\bm{\theta}}(\bm{s}(\bm{Y}))$.}

Let $p$ be the number of network terms, which describe the in-control state and can reflect the deviations {\color{blue}in the case of an} out-of-control state. Thus, {\color{blue} at time point $t$ there is a $p$-dimensional vector $\hat{\bm{c}}_t$ = $(\hat{c}_{1t}, \ldots, \hat{c}_{pt})'$ that estimates the network characteristics $\bm{c}$}.  Moreover, let $F_{\bm{c}_0, \bm{\Sigma}}$ be the target distribution of these estimates with $\bm{c}_{0} = \EX_0(\hat{c}_1, \ldots, \hat{c}_p)'$ being the expected value and $\bm{\Sigma}$ the respective $p \times p$ variance-covariance matrix {\color{blue} of the network characteristics} \citep{montgomery2012statistical}. 
Thus,
\begin{equation}\label{eq:cp_model}
\hat{\bm{c}}_t \quad {\sim} \quad \left\{ \begin{array}{cc}
F_{\bm{c}_0, \bm{\Sigma}} &  \text{ if } t < \tau \\
F_{\bm{c}_{\tau}, \bm{\Sigma}} & \text{ if } t \geq \tau \\
\end{array} \right. \, ,
\end{equation}
where $\tau$ denotes a change point to be detected and $\bm{c}_{\tau} \neq \bm{c}_0$. If $\tau = \infty$ the network is set to be in-control, whereas it is out of control in the case of $\tau \leq t < \infty$. Furthermore, we assume that the estimation precision of the parameters does not change across $t$, i.e., $\bm{\Sigma}$ is constant for the in-control and out-of-control state. Hence, the monitoring procedure is based on the expected values of $\hat{\bm{c}}_t$. In fact, we can specify the above mentioned hypothesis as follows
\begin{equation*}
H_{0, t}: \, \EX(\hat{\bm{c}}_t) = \bm{c}_{0} \qquad \text{against} \qquad H_{1, t}: \, \EX(\hat{\bm{c}}_t) \neq \bm{c}_{0} \, .
\end{equation*}

Typically, a multivariate control chart consists of the control statistic depending on one or more characteristic quantities, plotted in time order, and a horizontal line, called the upper control limit (UCL) that indicates the amount of acceptable variation. A hypothesis $H_0$ is rejected if the control statistic is equal to or exceeds the value of the UCL. 

Subsequently, we discuss several control statistics and present a method to determine the respective UCLs.

\subsection{Multivariate Cumulative Sum and Exponentially\\ Weighted Moving Average Control Charts}
The strength of the multivariate control chart over the univariate control chart is the ability to monitor several interrelated process variables. It implies that the corresponding test statistic should take into account the correlations of the data, be dimensionless and scale-invariant, as the process variables can considerably differ from each other. The squared Mahalanobis distance, which represents the general form of the control statistic, fulfils these criteria and is defined as
\begin{equation}
D_t^{(1)} = (\hat{\bm{c}}_t - \bm{c}_{0})' \bm{\Sigma}^{-1} (\hat{\bm{c}}_t - \bm{c}_{0}),
\end{equation}
being the part of the respective ``data depth'' expression -- Mahalanobis depth that measures a deviation from an in-control distribution (cf. \citealp{liu1995control}). Hence, $D_t^{(1)}$ maps the $p$-dimensional characteristic quantity $\hat{\bm{c}}_t$ to a one-dimensional measure. It is important to note that the characteristic quantity at time point $t$ is usually the mean of several samples at $t$, but in our case, we only observe one network at each instant of time. Thus, the characteristic quantity $\hat{\bm{c}}_t$ is the value of the obtained estimates and not the average of several samples.

{\color{blue} In this work, we apply two control chart types and compare their performance in network monitoring}. Firstly, we discuss multivariate CUSUM (MCUSUM) charts (cf. \citealp{woodall1985multivariate}; \citealp{pignatiello1990comparisons}; \citealp{ngai2001multivariate}). One of the widely used version was proposed by \cite{crosier1988multivariate} and is defined as follows \pagebreak
\begin{equation}
C_t = \big[(\bm{r}_{t-1} + \hat{\bm{c}}_t - \bm{c}_{0})'\bm{\Sigma}^{-1}(\bm{r}_{t-1} + \hat{\bm{c}}_t - \bm{c}_{0})\big]^{1/2},
\end{equation}
where
\begin{equation*}
\bm{r}_{t} = \begin{cases*}
\hspace*{3cm} \bm{0} \hspace*{3.9cm} \text{if}\ C_t \leq k, \\
\quad \quad \quad (\bm{r}_{t-1} + \hat{\bm{c}}_t - \bm{c}_{0})(1-k/C_t) \hspace*{1.4cm} \text{if}\ C_t > k, \\
\end{cases*}
\end{equation*}
given that $\bm{r}_0 = \bm{0}$ and $k>0$. The respective chart statistic is
\begin{equation}
D_t^{(2)} = \bm{r}'_t\bm{\Sigma}^{-1}\bm{r}_t,
\end{equation}
and it signals if $\sqrt{D_t^{(2)}}$ is greater than or equals the UCL. Certainly, the values $k$ and UCL considerably influence the performance of the chart. The parameter $k$, also known as reference value or allowance, reflects the variation tolerance, taking into consideration $\delta$ -- the deviation from the mean  measured in the standard deviation units we aim to detect. According to \cite{page1954continuous} and \cite{crosier1988multivariate},  the chart is approximately optimal if $k = \delta/2$.

Secondly, we consider multivariate charts based on exponential smoothing (EWMA). \cite{lowry1992multivariate} proposed a multivariate extension of the EWMA control chart (MEWMA), which is defined as follows
\begin{equation}
\bm{l}_t = \lambda(\hat{\bm{c}}_t - \bm{c}_0) + (1-\lambda)\bm{l}_{t-1}
\end{equation}
with the $0 < \lambda \leq 1$ and $\bm{l}_0 = \bm{0}$ (cf. \citealp{montgomery2012statistical}). The corresponding chart statistic is
\begin{equation}
D_t^{(3)} = \bm{l}'_t\bm{\Sigma}^{-1}_{\bm{l}_t}\bm{l}_t,
\end{equation}
where the covariance matrix is defined as
\begin{equation}
\bm{\Sigma}_{\bm{l}_t} = \dfrac{\lambda}{2-\lambda}\big[1-(1-\lambda)^{2t}\big]\bm{\Sigma}.
\end{equation}
Together with the MCUSUM, the MEWMA is an advisable approach for detecting relatively small but persistent changes. However, the detection of large shifts is also possible by setting $k$ or $\lambda$ high. For instance, in case of the MEWMA with $\lambda = 1$, the chart statistic coincides with $D_t^{(1)}$. Thus, it is equivalent to the Hotelling's $T^2$ control procedure, which is suitable for detection of substantial deviations. It is worth mentioning that the discussed methods are directionally invariant, therefore, the investigation of the data at the signal time point is necessary if the change direction is of particular interest.
\subsection{{\color{blue}Computation of Control Limits}}
Eventually, if $D_t^{(2)}$ or $D_t^{(3)}$ is equal to or exceeds the UCL, it means that the charts signal a change. To determine the UCLs, one typically assumes that the chart has a predefined (low) probability of false alarms, i.e., signals when the process is in control, or a prescribed in-control Average Run Length $ARL_0$, i.e., the number of expected time steps until the first signal. To compute the UCLs corresponding to $ARL_0$ {\color{blue} theoretically}, a prevalent number of multivariate control charts require a normally distributed target process (cf. \citealp{johnson2007applied, porzio2008multivariate, montgomery2012statistical}).  In our case, this assumption would need to be valid for the estimates of the model parameters{\color{blue}/the network statistics}. However, while there are some studies on the distributions of particular network statistics $\bm{s}(\bm{Y})$ (cf. \citealp{yan2013central, yan2016asymptotics, sambale2018logarithmic}), only a few results are obtained about the
parameter estimates of $\bm{\theta}$. Primarily, the difficulties to determine the distribution is that the assumption of independent and identically distributed data is violated in the ERGM case. In addition, the parameters depend on the choice of the model terms and network size \citep{he2015glmle}. \cite{kolaczyk2015question} proved asymptotic normality for the ML estimates in a simplified context of the ERGM, pointing out the necessity to establish a deeper understanding of the distributional properties of parameter estimates.

{\color{blue} In case that the normality assumption is violated to a slight or moderate degree, the control charts still will remain robust \citep{montgomery2012statistical}. The most crucial assumption that needs to be satisfied is the independence of the observations at different time points \citep{qiu2013introduction}. If the data is autocorrelated, the theoretically derived UCLs become invalid, so that their implementation would lead to inaccurate results. Here, we consider networks which are dependent over time. Moreover, the networks used for estimation of the characteristics $\hat{\bm{c}}_t$ are overlapping due to the application of the moving window approach. As shown in Section \ref{sec:ucls}, the characteristics that are based on the averaged network statistics $\hat{\bm{s}}_t$ can violate this assumption substantially. Regarding the estimates $\hat{\bm{\theta}}_t$, if their computation does not involve overlapping of the networks by the sliding window approach of size $z$, i.e., each graph is involved only once in the estimation of $\bm{\theta}$, and the size of $z$ is enough for recovering the temporal dependence completely,  then the estimates become independent. However, as we design an online monitoring procedure, we support the idea of estimating $\hat{\bm{c}}_t$ immediately as soon as a new data point is available. In this case, we account for the correlation between the estimated characteristics $\hat{\bm{c}}_t$.
	
	There are several works which apply control charts in presence of autocorrelation, advising either using the residuals of the time series models as observations, determining the theoretical control limits under autocorrelation or designing a simulation study for the estimation of the ARLs and corresponding control limits (cf. \citealp{montgomery1991some,  alwan1992effects, runger1995model, schmid1997some, zhang1997detection, lu1999control, lu2001cusum, sheu2009monitoring}). It is worth noting that the residual charts have different properties from the traditional charts, which we consider in this work. Hence, we determine the UCLs via Monte Carlo simulations described in Section \ref{sec:ucls}. }

\section{Simulation Study}\label{section:3}

To verify the applicability and effectiveness of the discussed approach, we design a simulation study followed by the surveillance of real-world data with the goal to obtain some insights into its temporal development.
%

\subsection{Estimation of the In-Control Parameters}

In practice, the in-control parameters $\bm{c}_{0}$ and $\bm{\Sigma}$ are usually unknown and therefore have to be estimated. Thus, one subdivides the sequence of network {\color{blue} observations} into Phase I and Phase II. In Phase I, the process must coincide with the in-control state {\color{blue} so that} the true in-control parameters $\bm{c}_{0}$ and $\bm{\Sigma}$ can be estimated by the sample mean vector $\bm{\bar{c}}$ and the sample covariance matrix $\bm{S}$ from  $\hat{\bm{c}}_t$. {\color{blue} This approach is particularly helpful because of the possibly substantial bias in the parameter
	estimates $\hat{\bm{\theta}}_t$ (cf. \citealp{van2009comparison}). Hence, the empirical determination of the target values allows us to not consider bias in our analysis. } 

It is important that Phase I replicates the natural behaviour of a network, so that if the network constantly grows, then it is vital to consider this aspect in Phase I. Similarly, if the type of network is prone to stay unchangeable in terms of additive connections or topological structure, this fact should be captured in Phase I for reliable estimation and later network surveillance. After the necessary estimates of $\bm{c}_{0}$, $\bm{\Sigma}$ and the UCL are obtained, the calibrated control chart can be applied to the actual data in Phase II.

\subsection{{\color{blue} Generation of Network Time Series}}\label{sec:networks}
To compute $\bm{\bar{c}}$ and $\bm{S}$, we need a certain number of in-control networks. For this purpose, we generate {\color{blue} 2500} temporal graphs of desired length $ T < \tau$, where each graph consists of $N = 100$ nodes. The simulation of synthetic networks is based on the Markov chain principle: {\color{blue}the network observation} in time point $\bm{Y}_t$ is simulated from {\color{blue}its previous state} $\bm{Y}_{t-1}$ by {\color{blue} selecting randomly a fraction $\phi$ of elements of the adjacency matrix and setting them} to either 1 or 0, according to a specified transition matrix $\bm{M}$. {\color{blue} This setting allows us to include the memory term during the estimation of the TERGM that reflects stability of both the edges and non-edges between the previous and the current network observation.} The in-control values are $\phi_0 = 0.01$ and

\begin{equation*}
\bm{M}_0 =
\begin{pmatrix}
m_{00,0} & m_{01,0} \\
m_{10,0} & m_{11,0}
\end{pmatrix} =
\begin{pmatrix}
0.9 & 0.1 \\
0.4 & 0.6
\end{pmatrix},
\end{equation*}
where $m_{ij,0}$ denotes the probability of a transition from $i$ to $j$ in the in-control state.

At the beginning of each sequence, a {\color{blue} directed }network which is called the ``base network'' is simulated by applying an ERGM with predefined network terms {\color{blue}and corresponding coefficients} so that it is possible to control the ``network creation'' indirectly. {\color{blue} This procedure helps to guarantee that the temporal networks have a stochastic but analogous initialisation.} In our case, we select three network statistics, namely an edge term, a triangle term and a parameter that defines asymmetric dyads. {\color{blue}These terms are used  later for estimating $\bm{c}_0$.} Subsequently, {\color{blue}a new graph is produced by applying the in-control fraction $\phi$ and the transition matrix $\bm{M}$}.

Next, we need to confirm that the generated samples of networks behave according to the requirements of Phase I, i.e., capturing only the usual variation of the target process. For this purpose, we can exploit Markov chain properties and calculate its steady-state equilibrium vector $\bm{\pi}$, as it follows that the expected number of non-edges and edges is given by $\bm{\pi}$. Using eigenvector decomposition, we find the steady-state to be $\bm{\pi} = (0.8, 0.2)'$. Consequently, the expected number of edges in the graph in its steady-state is 1980. However, the network density is only one of the aspects to define the in-control process, as the temporal development and the topology are also involved in the network creation. { \color{blue} Hence, we identify the suitable start of the considered network sequence by computing the network statistics $\bm{s}(\bm{Y}_t)$ over multiple network time series. By plotting the behaviour, we determined that all four terms become stable by $t=1000$.} Thus, we simulate {\color{blue}1000} network {\color{blue}observations} in a burn-in period so that the in-control {\color{blue}sequence} of network {\color{blue}states} starts at {\color{blue}$t = 1001$}.

\subsection{Calibration of the Charts {\color{blue}in Phase I}}\label{sec:ucls}
After the generation {\color{blue} of temporal networks, we compute $\hat{\bm{\theta}}_t$ by fitting the TERGM and $\hat{\bm{s}}_t$ by calculating the average with a certain window size $z$ using the four network terms, namely edge term, a triangle term, a term that defines asymmetric dyads and a memory} term which describes the stability of both edges and non-edges over time with {\color{blue}the temporal lag} $v=1$.   Currently, there are two widely used approaches {\color{blue} to estimate the TERGM}: Maximum Pseudolikelihood Estimation (MPLE) with bootstrapped confidence intervals and Markov Chain Monte Carlo (MCMC) ML estimation \citep{leifeld2018temporal}. The chosen estimation method to derive $\hat{\bm{\theta}}_t$ is the bootstrap MPLE which is appropriate to handle a relatively large number of nodes and time points \citep{leifeld2018temporal}. Next, we {\color{blue} calculate the in-control parameters $\bm{\bar{c}}$ and $\bm{S}$ for both monitoring cases. Finally, we calibrate different control charts by obtaining the UCLs with respect to the predefined $ARL_0$} via the bisection method. For two window sizes $z = \{7, 14\}$, Table \ref{MEWMA_UCL_newtheta} and \ref{MCUSUM_UCL_newtheta} {\color{blue}summarise the obtained results for surveillance of $\bm{\theta}$, and Table \ref{MEWMA_UCL_newstat} and \ref{MCUSUM_UCL_newstat} for surveillance of $\bm{s}(\bm{Y})$ with} the MEWMA and MCUSUM charts respectively. {\color{blue} If the reader wishes to apply the TERGM with the same network terms and similar window size as we did in this work, the presented UCLs can be used directly. Otherwise, it is necessary to conduct different Monte Carlo simulations that address the specific settings of the TERGM.}

{\color{blue}As both network characteristics describe the same process, we would expect the UCL results to be similar. However, in Figure \ref{ACF} the analysis of the autocorrelation functions applied to the estimates of one of the generated network time series shows that the dependence structures of $\hat{\bm{\theta}}_t$ and $\hat{\bm{s}}_t$ considerably differ.


\begin{table}[H]
	\small
	\begin{tabular}{llllllllllll}
		\toprule
		$z$&$\ ARL_0$/$\lambda$ &\makebox[3em]{0.1}&\makebox[2em]{0.2}&\makebox[2em]{0.3}&\makebox[2em]{0.4}
		&\makebox[2em]{0.5}&\makebox[2em]{0.6}&\makebox[2em]{0.7}&\makebox[2em]{0.8}&\makebox[2em]{0.9}&\makebox[2em]{1.0}\\
		\midrule
		&\makebox[4em]{50}&\makebox[3em]{39.32}&\makebox[2em]{35.76}&\makebox[2em]{31.04}&\makebox[2em]{26.52}
		&\makebox[2em]{22.58}&\makebox[2em]{19.22}&\makebox[2em]{16.46}&\makebox[2em]{14.10}&\makebox[2em]{12.14}&\makebox[2em]{10.46}\\
		7 & \makebox[4em]{75} &\makebox[3em]{45.76}&\makebox[2em]{41.03}&\makebox[2em]{35.20}&\makebox[2em]{29.81}
		&\makebox[2em]{25.27}&\makebox[2em]{21.61}&\makebox[2em]{18.53}&\makebox[2em]{15.86}&\makebox[2em]{13.66}&\makebox[2em]{11.72}\\
		& \makebox[4em]{100} &\makebox[3em]{50.52}&\makebox[2em]{45.30}&\makebox[2em]{38.58}&\makebox[2em]{32.56}
		&\makebox[2em]{27.62}&\makebox[2em]{23.48}&\makebox[2em]{20.00}&\makebox[2em]{17.13}&\makebox[2em]{14.71}&\makebox[2em]{12.66}\\
		
		\midrule
		
		&\makebox[4em]{50}&\makebox[3em]{55.14}&\makebox[2em]{43.44}&\makebox[2em]{33.80}&\makebox[2em]{26.96}
		&\makebox[2em]{21.98}&\makebox[2em]{18.16}&\makebox[2em]{15.16}&\makebox[2em]{12.76}&\makebox[2em]{10.76}&\makebox[2em]{9.15}\\
		14 & \makebox[4em]{75}&\makebox[3em]{65.63}&\makebox[2em]{50.94}&\makebox[2em]{39.29}&\makebox[2em]{31.23}
		&\makebox[2em]{25.26}&\makebox[2em]{20.68}&\makebox[2em]{17.24}&\makebox[2em]{14.50}&\makebox[2em]{12.20}&\makebox[2em]{10.32}\\
		& \makebox[4em]{100} &\makebox[3em]{73.85}&\makebox[2em]{56.20}&\makebox[2em]{42.97}&\makebox[2em]{33.97}
		&\makebox[2em]{27.44}&\makebox[2em]{22.52}&\makebox[2em]{18.72}&\makebox[2em]{15.69}&\makebox[2em]{13.21}&\makebox[2em]{11.15}\\
		
		\bottomrule
	\end{tabular}
	\caption{ {\color{blue}Upper control limits for the MEWMA chart based on the estimates $\hat{\bm{\theta}}_t$ and \\$ARL_0 \in \{50, 75, 100\}$ for two different windows sizes $z = 7$ and $z = 14$. }}
	\label{MEWMA_UCL_newtheta}
\end{table}

\begin{table}[H]
	\small
	\begin{tabular}{llllllllllll}
		\toprule
		$z$&$\ ARL_0$/$\lambda$ &\makebox[3em]{0.1}&\makebox[2em]{0.2}&\makebox[2em]{0.3}&\makebox[2em]{0.4}
		&\makebox[2em]{0.5}&\makebox[2em]{0.6}&\makebox[2em]{0.7}&\makebox[2em]{0.8}&\makebox[2em]{0.9}&\makebox[2em]{1.0}\\
		\midrule
		&\makebox[4em]{50}&\makebox[3em]{65.03}&\makebox[2em]{43.79}&\makebox[2em]{32.23}&\makebox[2em]{25.29}
		&\makebox[2em]{20.36}&\makebox[2em]{16.72}&\makebox[2em]{13.91}&\makebox[2em]{11.66}&\makebox[2em]{9.85}&\makebox[2em]{8.31}\\
		7 & \makebox[4em]{75} &\makebox[3em]{82.40}&\makebox[2em]{52.41}&\makebox[2em]{38.13}&\makebox[2em]{29.52}
		&\makebox[2em]{23.59}&\makebox[2em]{19.23}&\makebox[2em]{15.88}&\makebox[2em]{13.23}&\makebox[2em]{11.13}&\makebox[2em]{9.42}\\
		& \makebox[4em]{100} &\makebox[3em]{96.23}&\makebox[2em]{59.39}&\makebox[2em]{42.92}&\makebox[2em]{32.96}
		&\makebox[2em]{26.19}&\makebox[2em]{21.24}&\makebox[2em]{17.58}&\makebox[2em]{14.69}&\makebox[2em]{12.32}&\makebox[2em]{10.35}\\
		
		\midrule
		
		&\makebox[4em]{50}&\makebox[3em]{71.09}&\makebox[2em]{45.47}&\makebox[2em]{32.66}&\makebox[2em]{24.81}
		&\makebox[2em]{19.51}&\makebox[2em]{15.80}&\makebox[2em]{12.95}&\makebox[2em]{10.74}&\makebox[2em]{8.96}&\makebox[2em]{7.53}\\
		14 & \makebox[4em]{75}&\makebox[3em]{89.03}&\makebox[2em]{55.26}&\makebox[2em]{38.71}&\makebox[2em]{29.06}
		&\makebox[2em]{22.85}&\makebox[2em]{18.40}&\makebox[2em]{15.07}&\makebox[2em]{12.46}&\makebox[2em]{10.35}&\makebox[2em]{8.66}\\
		& \makebox[4em]{100} &\makebox[3em]{103.00}&\makebox[2em]{62.73}&\makebox[2em]{43.73}&\makebox[2em]{32.56}
		&\makebox[2em]{25.40}&\makebox[2em]{20.40}&\makebox[2em]{16.65}&\makebox[2em]{13.73}&\makebox[2em]{11.43}&\makebox[2em]{9.57}\\
		\bottomrule

	\end{tabular}
	\caption{ {\color{blue}Upper control limits for the MEWMA chart based on the estimates $\hat{\bm{s}}_t$ and\\ $ARL_0 \in \{50, 75, 100\}$ for two different windows sizes $z = 7$ and $z = 14$.}}
	\label{MEWMA_UCL_newstat}
\end{table}

\begin{table}[H]
	\small
	\begin{tabular}{lllllllllllll}
		\toprule
		$z$&$\ ARL_0$/$k$ &\makebox[3em]{0.5}&\makebox[2em]{0.6}&\makebox[2em]{0.7}&\makebox[2em]{0.8}
		&\makebox[2em]{0.9}&\makebox[2em]{1.0}&\makebox[2em]{1.1}&\makebox[2em]{1.2}&\makebox[2em]{1.3}&\makebox[2em]{1.4}&\makebox[2em]{1.5}\\
		\midrule
		&\makebox[4em]{50}&\makebox[3em]{21.01}&\makebox[2em]{19.36}&\makebox[2em]{17.75}&\makebox[2em]{16.21}
		&\makebox[2em]{14.83}&\makebox[2em]{13.52}&\makebox[2em]{12.25}&\makebox[2em]{11.10}&\makebox[2em]{9.99}&\makebox[2em]{9.00}&\makebox[2em]{8.03}\\
		7 & \makebox[4em]{75} &\makebox[3em]{25.19}&\makebox[2em]{22.90}&\makebox[2em]{20.82}&\makebox[2em]{18.95}
		&\makebox[2em]{17.33}&\makebox[2em]{15.91}&\makebox[2em]{14.47}&\makebox[2em]{13.19}&\makebox[2em]{11.94}&\makebox[2em]{10.74}&\makebox[2em]{9.61}\\
		& \makebox[4em]{100} &\makebox[3em]{28.25}&\makebox[2em]{25.59}&\makebox[2em]{23.31}&\makebox[2em]{21.18}&\makebox[2em]{19.38}
		&\makebox[2em]{17.67}&\makebox[2em]{16.12}&\makebox[2em]{14.67}&\makebox[2em]{13.27}&\makebox[2em]{11.96}&\makebox[2em]{10.73}\\
		\midrule
		
		&\makebox[4em]{50}&\makebox[3em]{30.10}&\makebox[2em]{27.84}&\makebox[2em]{25.64}&\makebox[2em]{23.67}
		&\makebox[2em]{21.69}&\makebox[2em]{19.83}&\makebox[2em]{17.92}&\makebox[2em]{16.17}&\makebox[2em]{14.44}&\makebox[2em]{12.75}&\makebox[2em]{11.06}\\
		14 & \makebox[4em]{75}&\makebox[3em]{37.35}&\makebox[2em]{34.60}&\makebox[2em]{31.91}&\makebox[2em]{29.25}
		&\makebox[2em]{26.86}&\makebox[2em]{24.53}&\makebox[2em]{22.37}&\makebox[2em]{20.20}&\makebox[2em]{18.14}&\makebox[2em]{16.19}&\makebox[2em]{14.32}\\
		& \makebox[4em]{100}&\makebox[3em]{43.06}&\makebox[2em]{39.52}&\makebox[2em]{36.20}&\makebox[2em]{33.15}
		&\makebox[2em]{30.45}&\makebox[2em]{27.84}&\makebox[2em]{25.43}&\makebox[2em]{23.16}&\makebox[2em]{20.97}&\makebox[2em]{18.81}&\makebox[2em]{16.77}\\
		\bottomrule
	\end{tabular}
	\caption{{\color{blue}Upper control limits for the MCUSUM chart based on the estimates $\hat{\bm{\theta}}_t$ and \\ $ARL_0 \in \{50, 75, 100\}$ for two different windows sizes $z = 7$ and $z = 14$.}}
	\label{MCUSUM_UCL_newtheta}
\end{table}

\begin{table}[H]
	\small
	\begin{tabular}{lllllllllllll}
		\toprule
		$z$&$\ ARL_0$/$k$ &\makebox[3em]{0.5}&\makebox[2em]{0.6}&\makebox[2em]{0.7}&\makebox[2em]{0.8}
		&\makebox[2em]{0.9}&\makebox[2em]{1.0}&\makebox[2em]{1.1}&\makebox[2em]{1.2}&\makebox[2em]{1.3}&\makebox[2em]{1.4}&\makebox[2em]{1.5}\\
		\midrule
		&\makebox[4em]{50}&\makebox[3em]{51.85}&\makebox[2em]{47.41}&\makebox[2em]{42.96}&\makebox[2em]{38.27}
		&\makebox[2em]{33.87}&\makebox[2em]{29.71}&\makebox[2em]{25.94}&\makebox[2em]{22.51}&\makebox[2em]{19.01}&\makebox[2em]{15.79}&\makebox[2em]{13.06}\\
		7 & \makebox[4em]{75} &\makebox[3em]{75.93}&\makebox[2em]{69.26}&\makebox[2em]{62.97}&\makebox[2em]{56.83}
		&\makebox[2em]{50.79}&\makebox[2em]{44.82}&\makebox[2em]{39.02}&\makebox[2em]{33.73}&\makebox[2em]{29.05}&\makebox[2em]{24.95}&\makebox[2em]{20.95}\\
		& \makebox[4em]{100} &\makebox[3em]{97.58}&\makebox[2em]{89.46}&\makebox[2em]{81.68}&\makebox[2em]{73.51}
		&\makebox[2em]{65.78}&\makebox[2em]{59.01}&\makebox[2em]{52.43}&\makebox[2em]{45.96}&\makebox[2em]{39.96}&\makebox[2em]{34.40}&\makebox[2em]{29.23}\\
		\midrule
		
		&\makebox[4em]{50}&\makebox[3em]{55.72}&\makebox[2em]{51.27}&\makebox[2em]{46.63}&\makebox[2em]{41.90}
		&\makebox[2em]{37.54}&\makebox[2em]{33.29}&\makebox[2em]{29.18}&\makebox[2em]{25.46}&\makebox[2em]{21.69}&\makebox[2em]{18.21}&\makebox[2em]{15.36}\\
		14 & \makebox[4em]{75}&\makebox[3em]{80.28}&\makebox[2em]{73.70}&\makebox[2em]{67.32}&\makebox[2em]{61.13}
		&\makebox[2em]{54.75}&\makebox[2em]{48.86}&\makebox[2em]{43.15}&\makebox[2em]{37.76}&\makebox[2em]{32.81}&\makebox[2em]{28.26}&\makebox[2em]{24.20}\\
		& \makebox[4em]{100}&\makebox[3em]{102.34}&\makebox[2em]{94.25}&\makebox[2em]{85.88}&\makebox[2em]{78.05}
		&\makebox[2em]{70.90}&\makebox[2em]{63.96}&\makebox[2em]{57.03}&\makebox[2em]{50.65}&\makebox[2em]{44.31}&\makebox[2em]{38.71}&\makebox[2em]{33.39}\\
		\bottomrule
	\end{tabular}
	\caption{{\color{blue}Upper control limits for the MCUSUM chart based on the estimates $\hat{\bm{s}}_t$ and\\ $ARL_0 \in \{50, 75, 100\}$ for two different windows sizes $z = 7$ and $z = 14$.}}
	\label{MCUSUM_UCL_newstat}
\end{table}

\begin{figure}[H]
	\centering
	\begin{subfigure}[b]{\textwidth}
		\centering
		\includegraphics[width=0.48\linewidth]{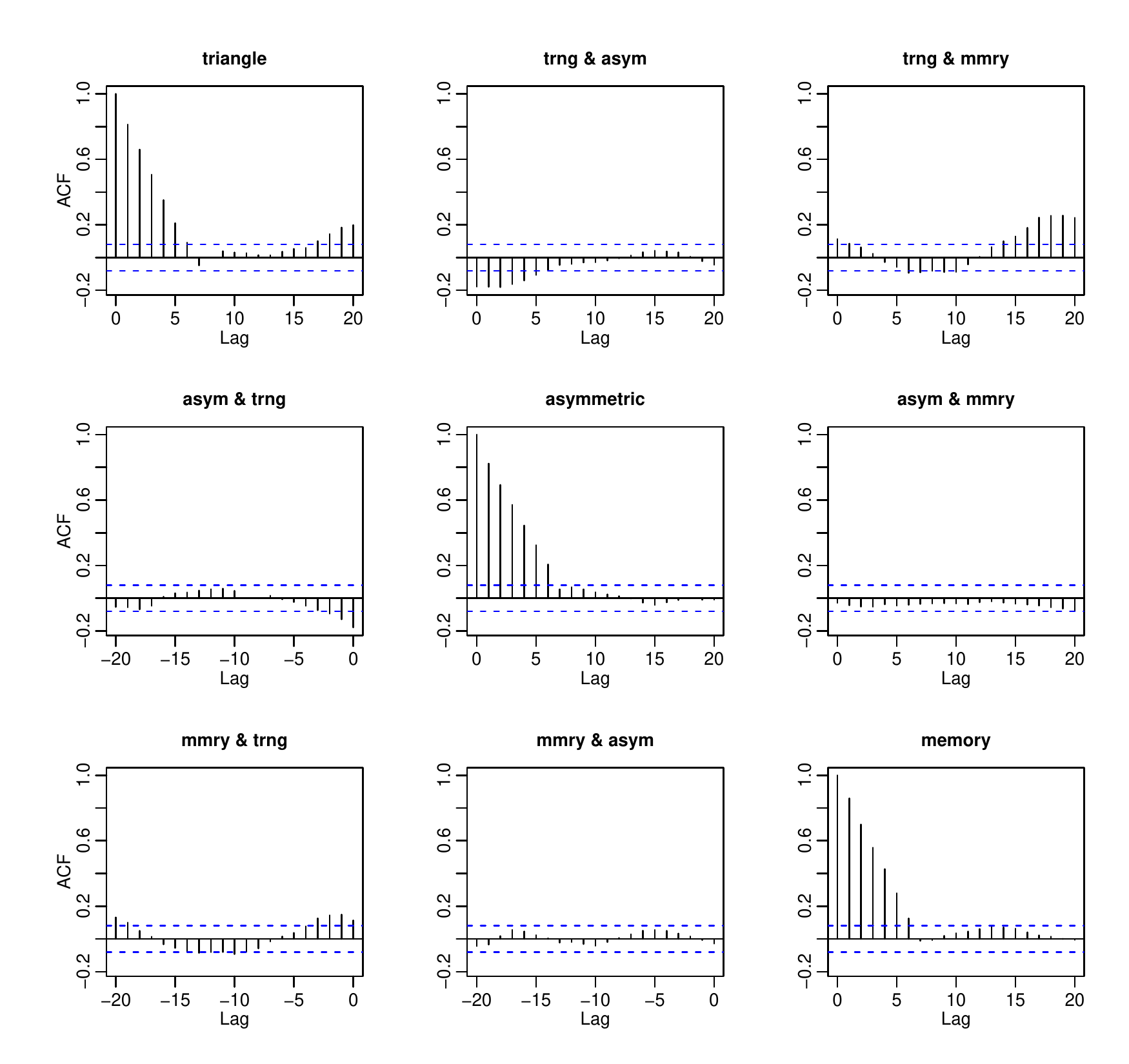}%
		\hfill
		\includegraphics[width=0.48\linewidth]{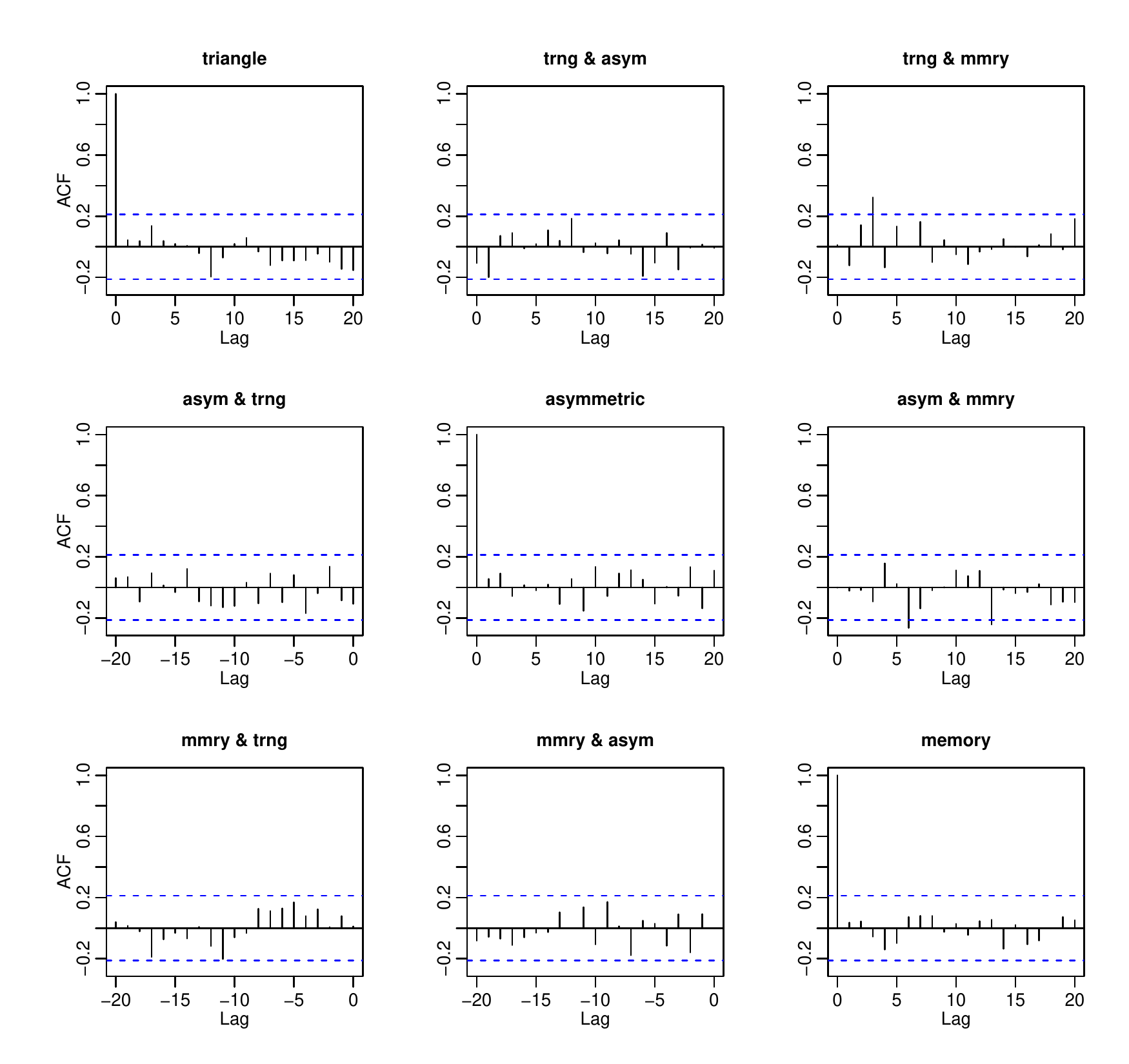}
		\caption{ACF of the estimates $\hat{\bm{\theta}}_t$ on the example of triangle, asymmetric and memory network terms.}
	\end{subfigure}
	\vskip\baselineskip
	\begin{subfigure}[b]{\textwidth}
		\centering
		\includegraphics[width=0.48\linewidth]{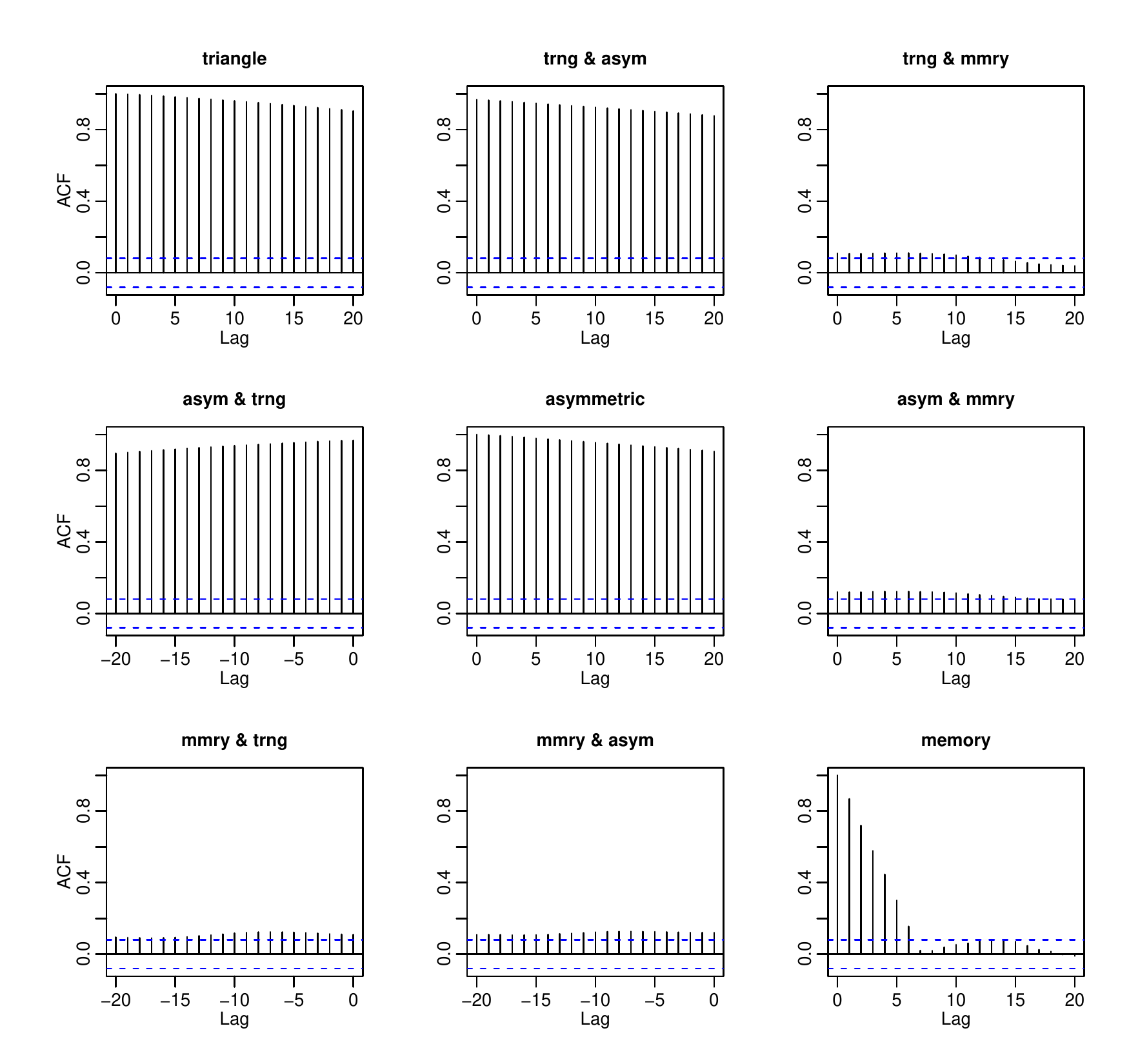}%
		\hfill
		\includegraphics[width=0.48\linewidth]{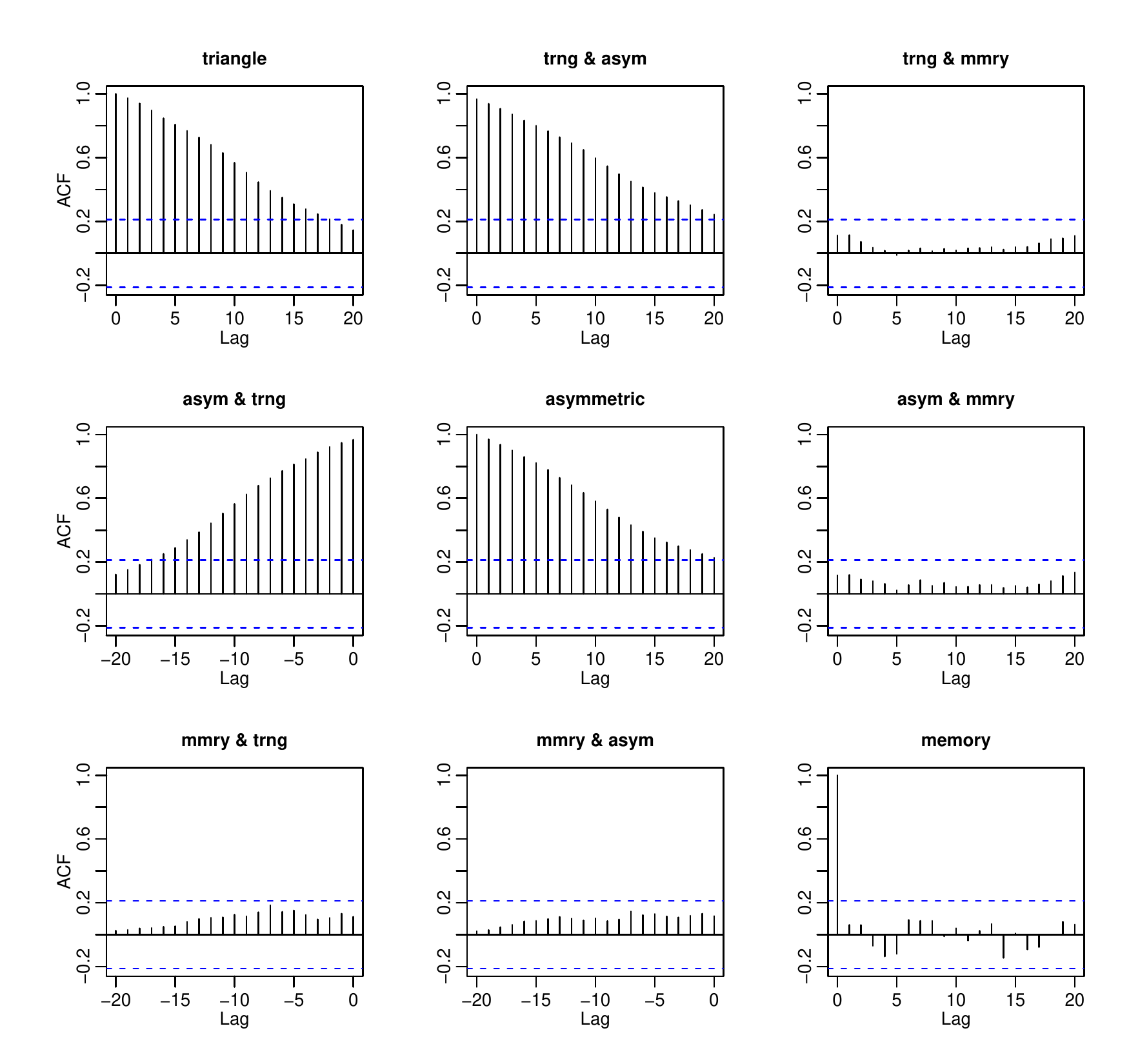}
		\caption{ACF of the estimates $\hat{\bm{s}}_t$ on the example of triangle, asymmetric and memory network terms.}
	\end{subfigure}
	\caption{{\color{blue}Comparison of the autocorrelation function (ACF) values when the network characteristics are estimated with a sliding window approach of size $z = 7$ containing (left) and not containing (right) overlapping network states.}}
	\label{ACF}
\end{figure}
\noindent While the elimination of the overlap in the calculation procedure removes correlation in the case of the parameter estimates $\hat{\bm{\theta}}_t$, there is only a slight improvement regarding the averaged network statistics $\hat{\bm{s}}_t$.   Thus, the UCLs are different for both cases.

}


\subsection{{\color{blue}Design of the Anomalous Behaviour}}\label{sec:anomalies}

\begin{figure}[H]
	\begin{center}
		\includegraphics[width=0.9\textwidth, trim= 0 0 0 0,clip]{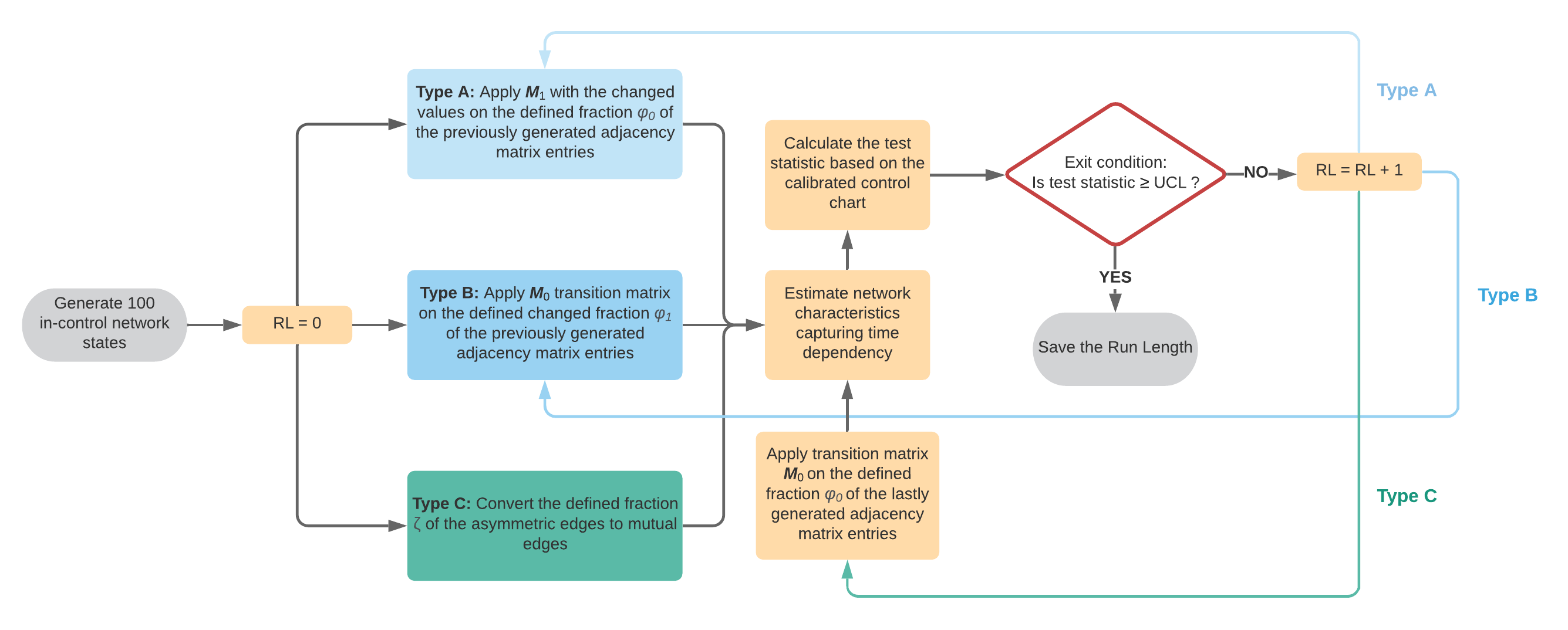}
		\caption{{\color{blue}Anomaly types for the generation of observations from Phase II and calculation of the associated run length.}}
		\label{fig:PhaseII}
	\end{center}
\end{figure}

{\color{blue} To test how well the proposed control charts can detect the changes in the network's development, it is necessary to compose different anomalous cases and generate samples from Phase II.} {\color{blue}Since our focus is on the detection of shifts in the process mean}, {\color{blue} an anomalous change can occur  {\color{blue}  either in the proportion of the asymmetric edges, in} the fraction of the randomly selected adjacency matrix entries $\phi$  or} the transition matrix $\bm{M}$. Thus, we subdivide these scenarios into three different anomaly types which are briefly described in the flow chart presented in Figure \ref{fig:PhaseII}.

We define a Type A anomaly as a change in the values of $\bm{M}$. That is, there is a transition matrix $\bm{M}_1 \neq \bm{M}_0$ when $t \geq \tau$. {\color{blue} Similar to Type A,} we consider anomalies of Type B by introducing a new fraction value $\phi_1$ in the generation process when $t \geq \tau$.  {\color{blue}Both types are instances of a persistent change (also known as simply a ``change''), where the abnormal development continues for all $t \geq \tau$ \citep{ranshous_shen_koutra_harenberg_faloutsos_samatova_2015}.} Anomalies of Type C differ from the previous two types as they represent a ``point change'' {\color{blue}(also referred to as an ``event'')}  -- the abnormal behaviour occurs only at a single point of time {\color{blue}$\tau$} but its outcome may also affect subsequent network states in our case {\color{blue} due to the Markov property}. We recreate this type of anomaly by converting a fraction $\zeta$ of asymmetric edges into mutual links. This process happens at time point $\tau$ only. Afterwards, the new network states are created similar to Phase I by applying $\bm{M}_0$ and $\phi_0$ up until the anomaly is detected. The considered cases are summarised in Table \ref{Anomalies}.

\begin{table}
	\centering
	\small
	\begin{tabular}{|c|c|c|c|}
		
		\hline
		\multicolumn{1}{|c|}{Anomaly Type} &
		\multicolumn{1}{c|}{Description} &
		\multicolumn{2}{c|}{Case} \\
		\hline

		Type A & Change in the transition matrix $\bm{M}$      & A.1 &  \makecell{ $m_{00,1} = 0.89$ ($m_{00,0} = 0.9$)  \\
			$m_{01,1} = 0.11$ ($m_{01,0} = 0.1$)} \\ \cline{3-4}
		&                                               & A.2 &  \makecell{ $m_{10,1} = 0.6$  ($m_{10,0} = 0.4$)  \\
			$m_{11,1} = 0.4$  ($m_{11,0} = 0.6$)} \\ \cline{3-4}
		&                                               & A.3 &  \makecell{ $m_{00,1} = 0.5$  ($m_{00,0} = 0.9$)  \\
			$m_{11,1} = 0.5$  ($m_{11,0} = 0.6$) }  \\ \cline{3-4}

		\hline
		\hline
		Type B & Change of the fraction $\phi_0=0.01$                         & B.1 & $\phi_1 = 0.009$  \\ \cline{3-4}
		&                                                 & B.2 & $\phi_1 = 0.015$  \\ \cline{3-4}
		&                                                 & B.3 & $\phi_1 = 0.02$   \\
		\hline
		\hline
		Type C & Increase of the proportion        & C.1 & $\zeta = 0.005$  \\ \cline{3-4}
		& of mutual edges by $\zeta$          & C.2 & $\zeta = 0.01$   \\ \cline{3-4}
		&                                                 & C.3 & $\zeta = 0.05$   \\
		
		\hline
	\end{tabular}
	\caption[types]{Anomaly cases. }
	\label{Anomalies}
\end{table}

\subsection{Performance of the Charts {\color{blue}in Phase II}}\label{sec:ced_study}
\begin{figure}[H]
	\centering
	\subfloat[MCUSUM, Case A.1]{\includegraphics[width=0.5\textwidth, trim = 0cm 0.5cm 0cm 2cm, clip]{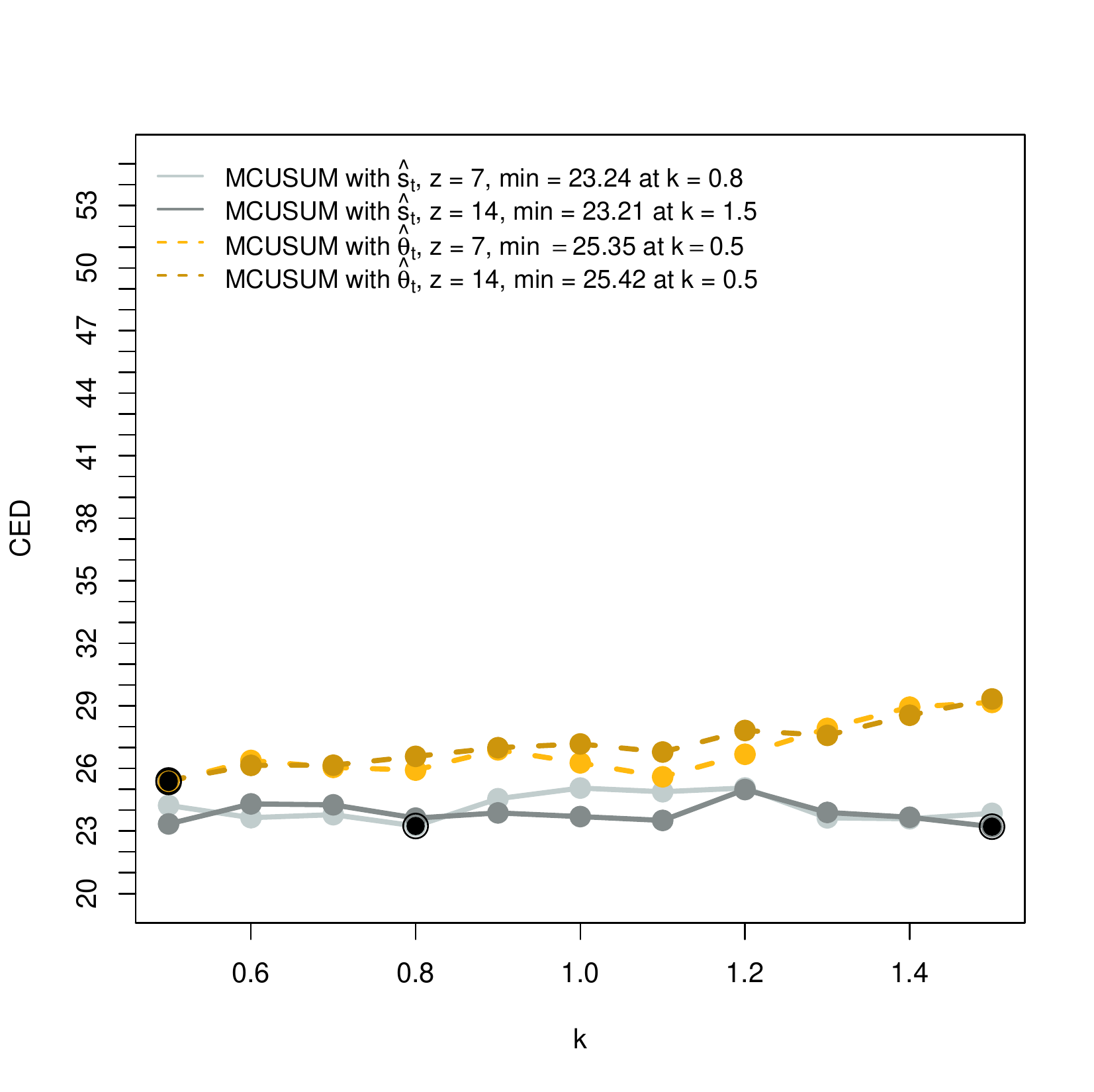}}
	\hfill
	\subfloat[MEWMA, Case A.1]{\includegraphics[width=0.5\textwidth, trim = 0cm 0.5cm 0cm 2cm, clip]{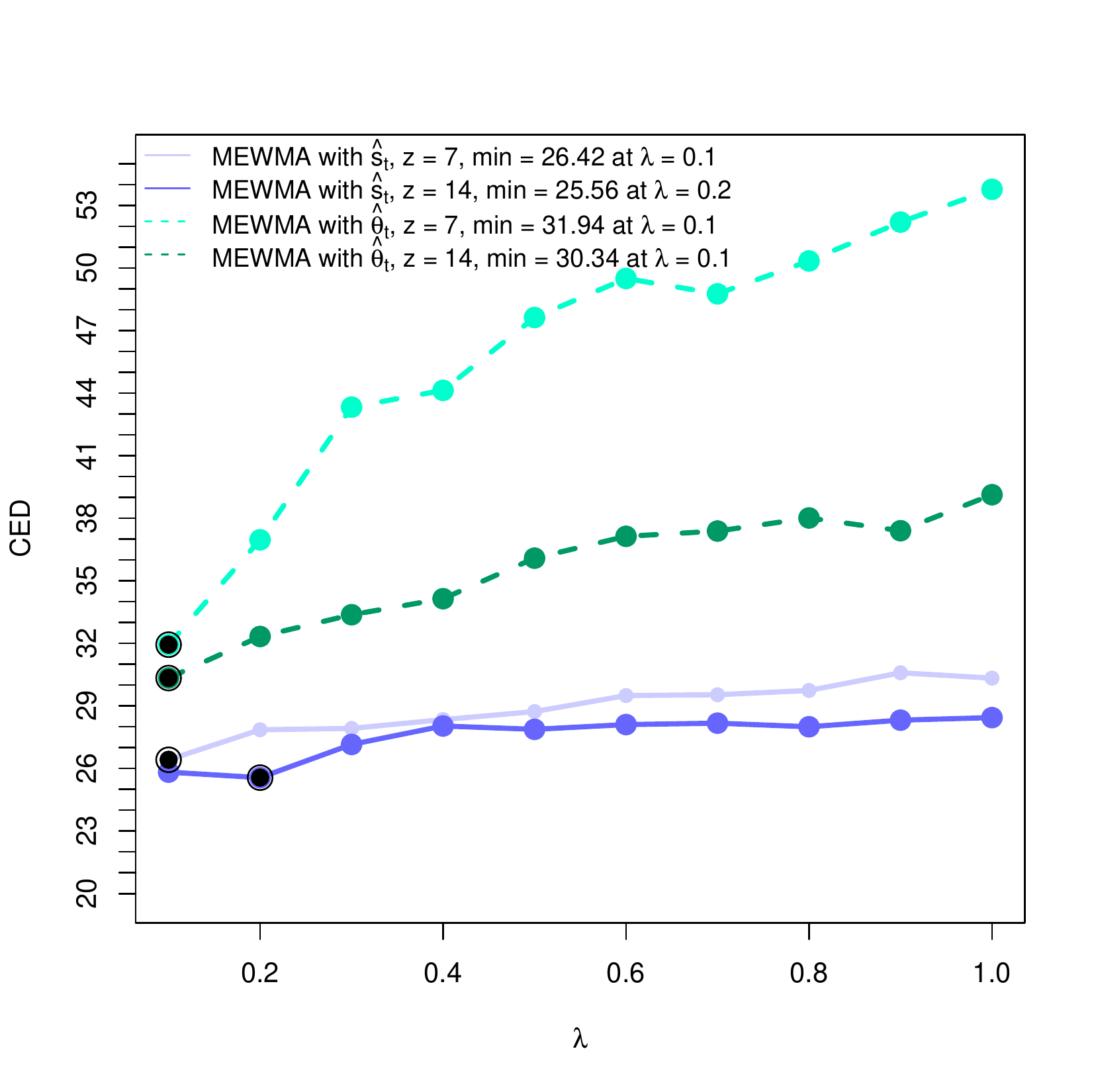}}
	\hfill
	\subfloat[MCUSUM, Case A.2]{\includegraphics[width=0.5\textwidth, trim = 0cm 0.5cm 0cm 2cm, clip]{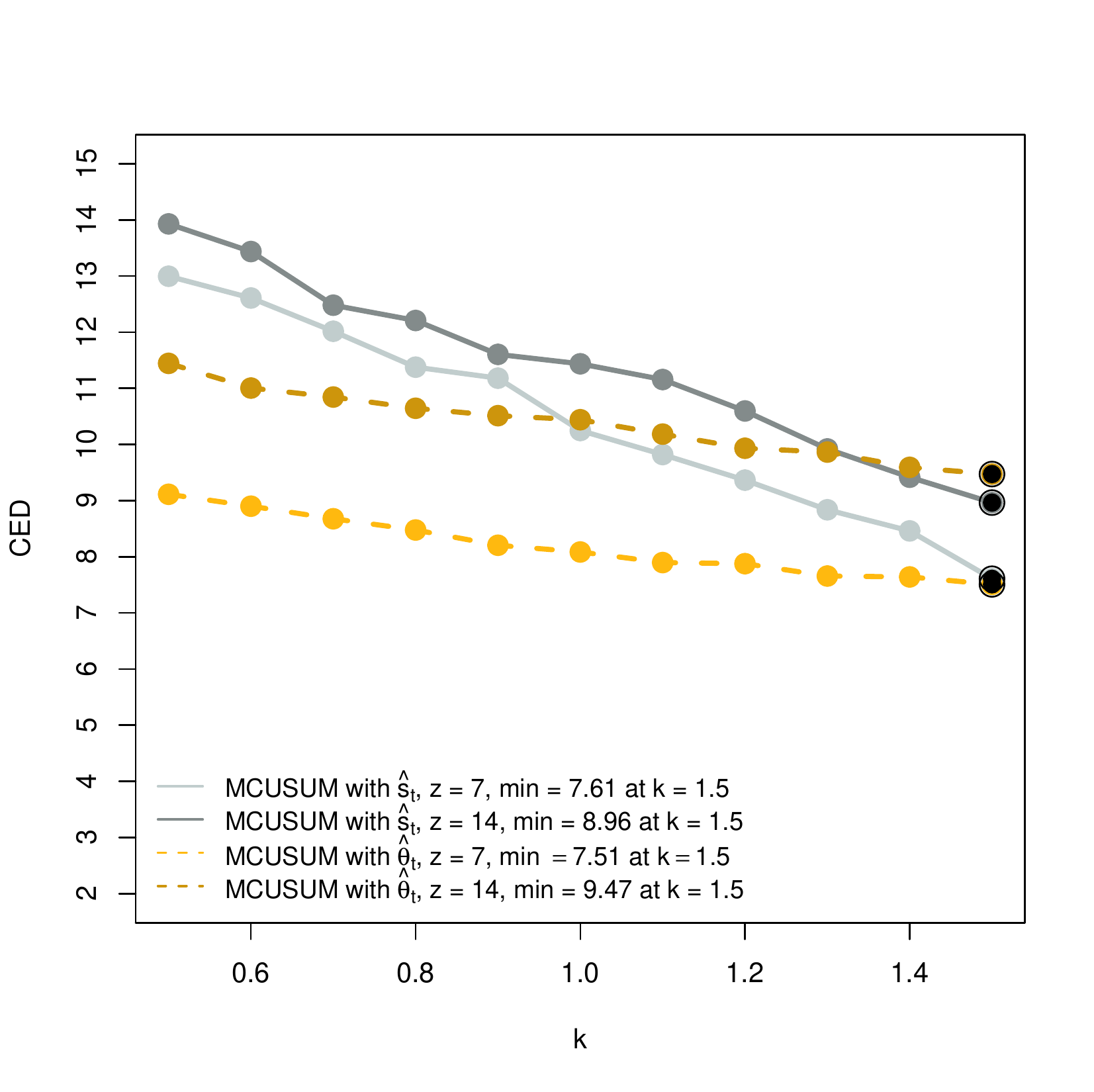}}
	\hfill
	\subfloat[MEWMA, Case A.2]{\includegraphics[width=0.5\textwidth, trim = 0cm 0.5cm 0cm 2cm, clip]{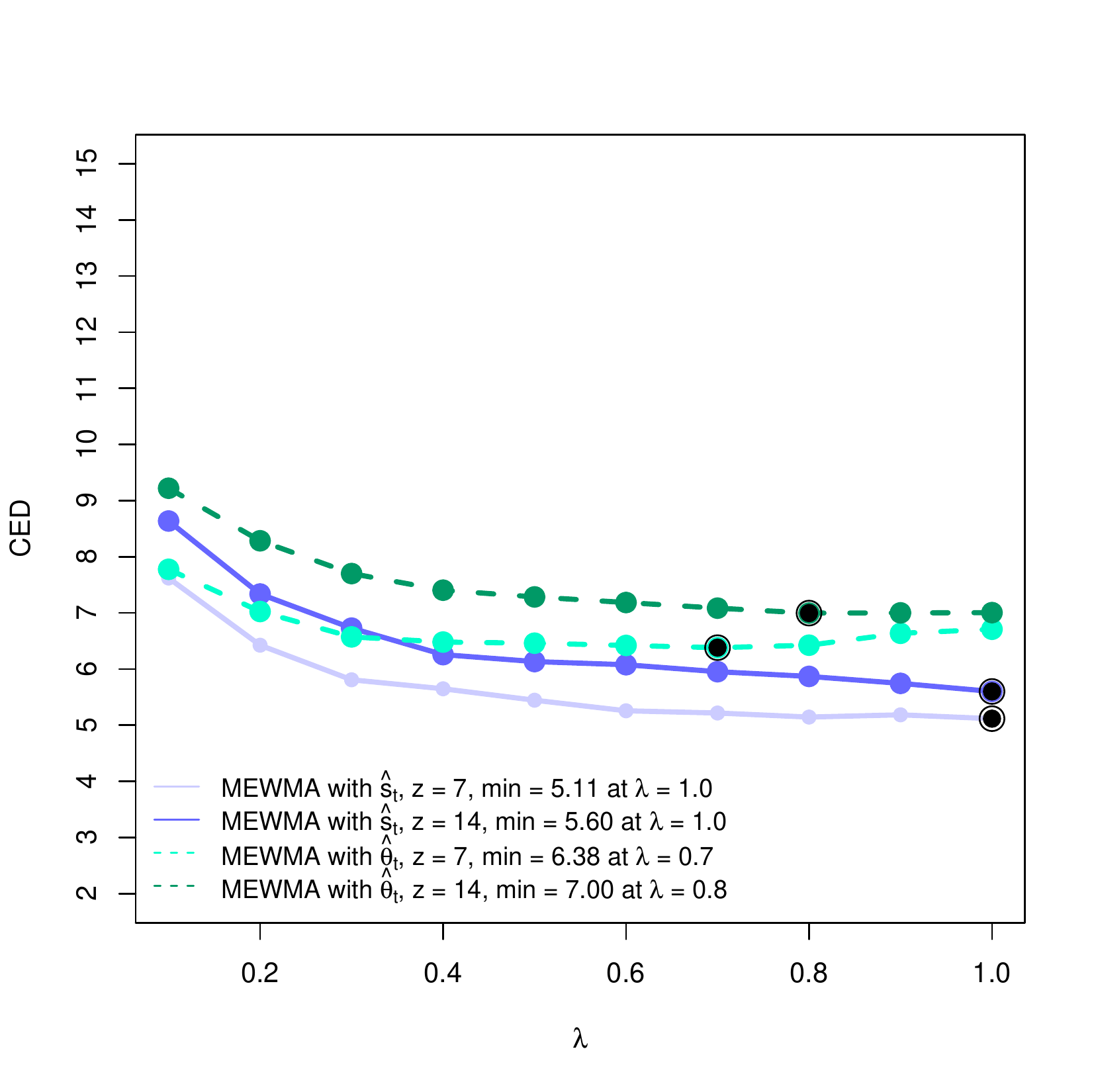}}
	\hfill
	\subfloat[MCUSUM, Case A.3]{\includegraphics[width=0.5\textwidth, trim = 0cm 0.5cm 0cm 2cm, clip]{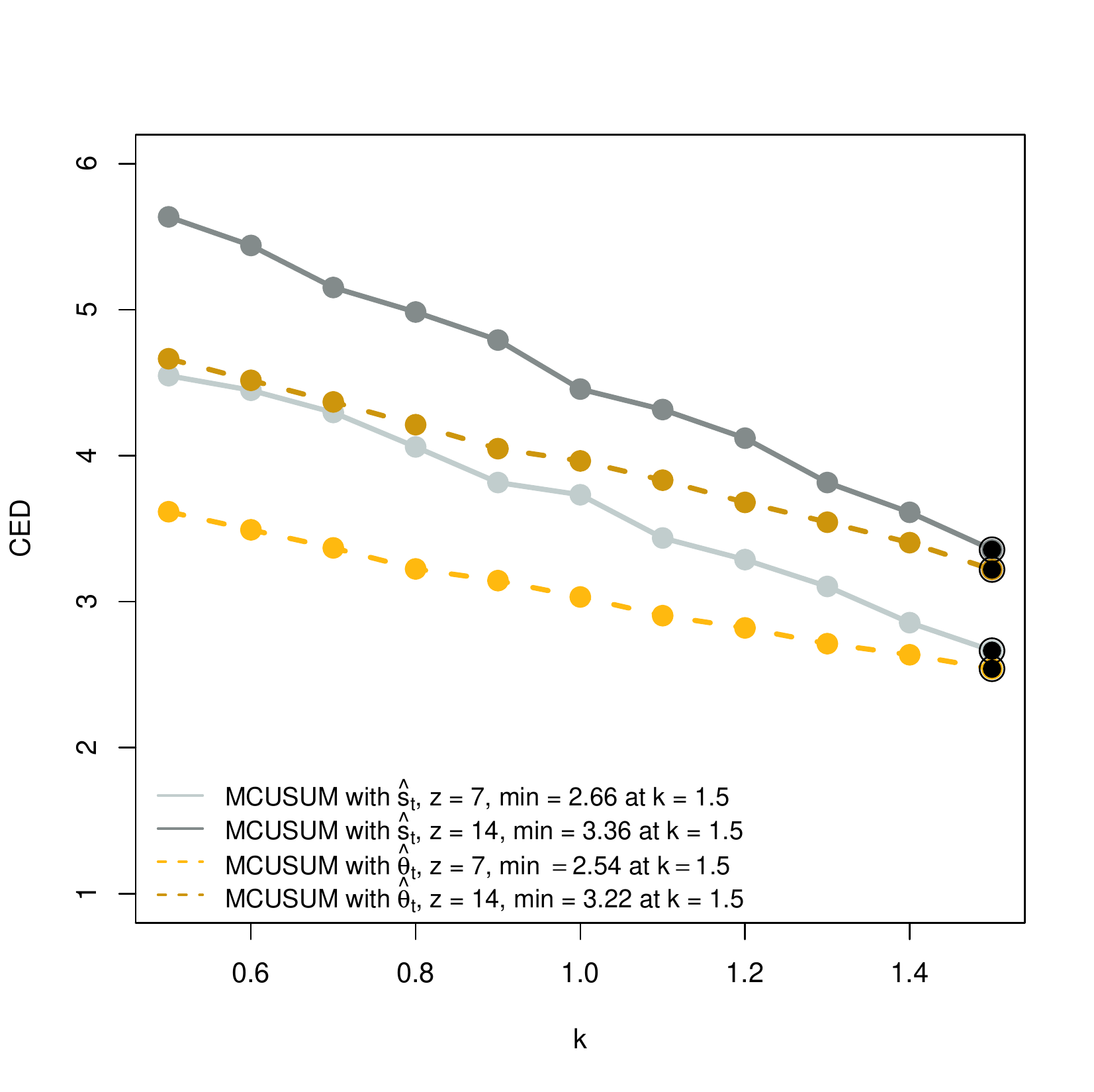}}
	\hfill
	\subfloat[MEWMA, Case A.3]{\includegraphics[width=0.5\textwidth, trim = 0cm 0.5cm 0cm 2cm, clip]{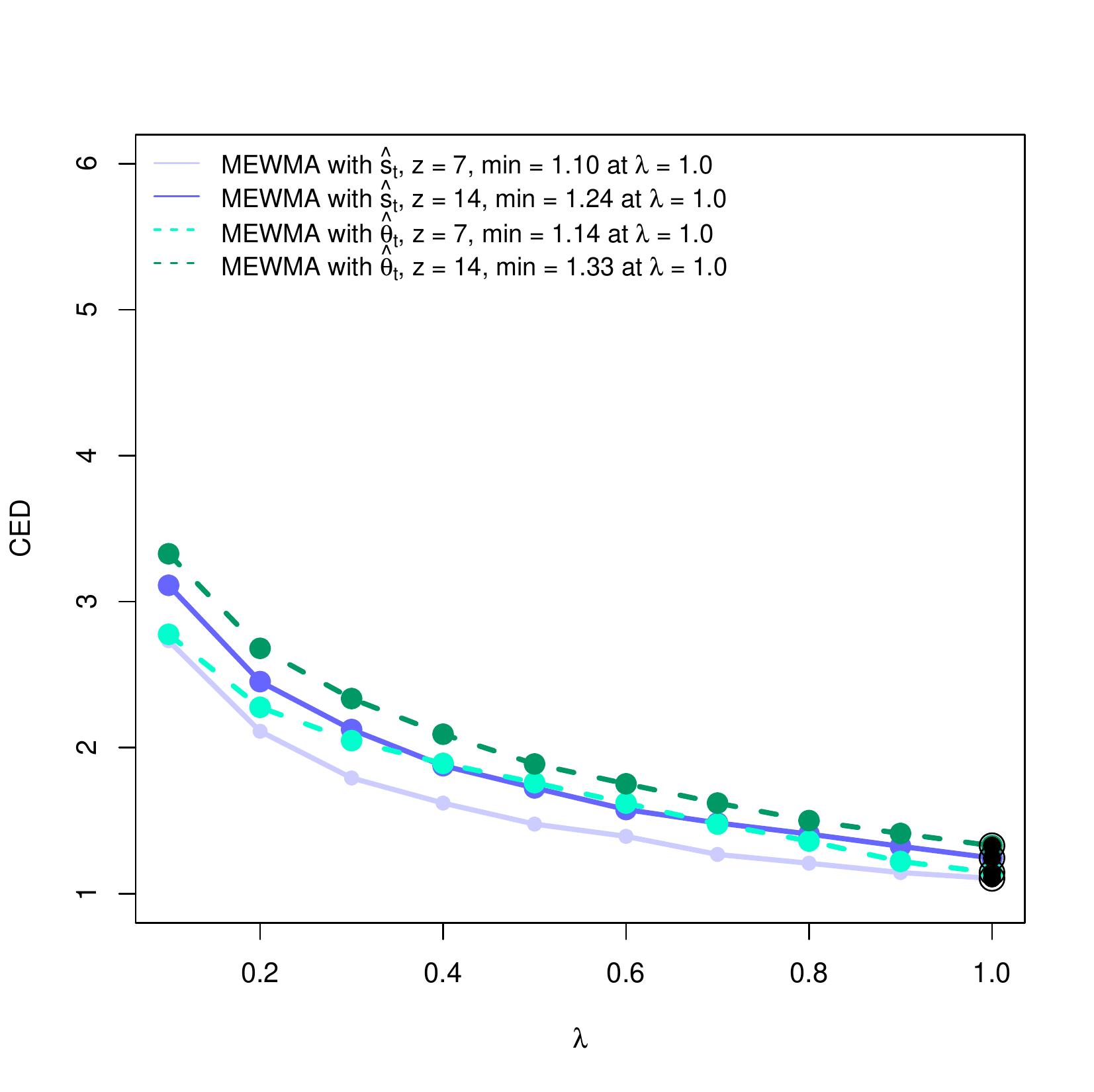}}
	
	\caption{{\color{blue}Conditional expected delays for anomalies of Type A for MCUSUM (left) and MEWMA (right) together with the different choices of the reference parameter $k$ and the smoothing parameter $\lambda$, the window sizes $z = 7$ and $z = 14$, and the network estimates $\hat{\bm{s}}_t$ (solid lines) and $\hat{\bm{\theta}}_t$ (dashed lines). Black points indicate the minimum CED for each setting.}}
	\label{MType1}
\end{figure}

\begin{figure}[H]
	\centering
	\subfloat[MCUSUM, Case B.1]{\includegraphics[width=0.5\textwidth, trim = 0cm 0.5cm 0cm 2cm, clip]{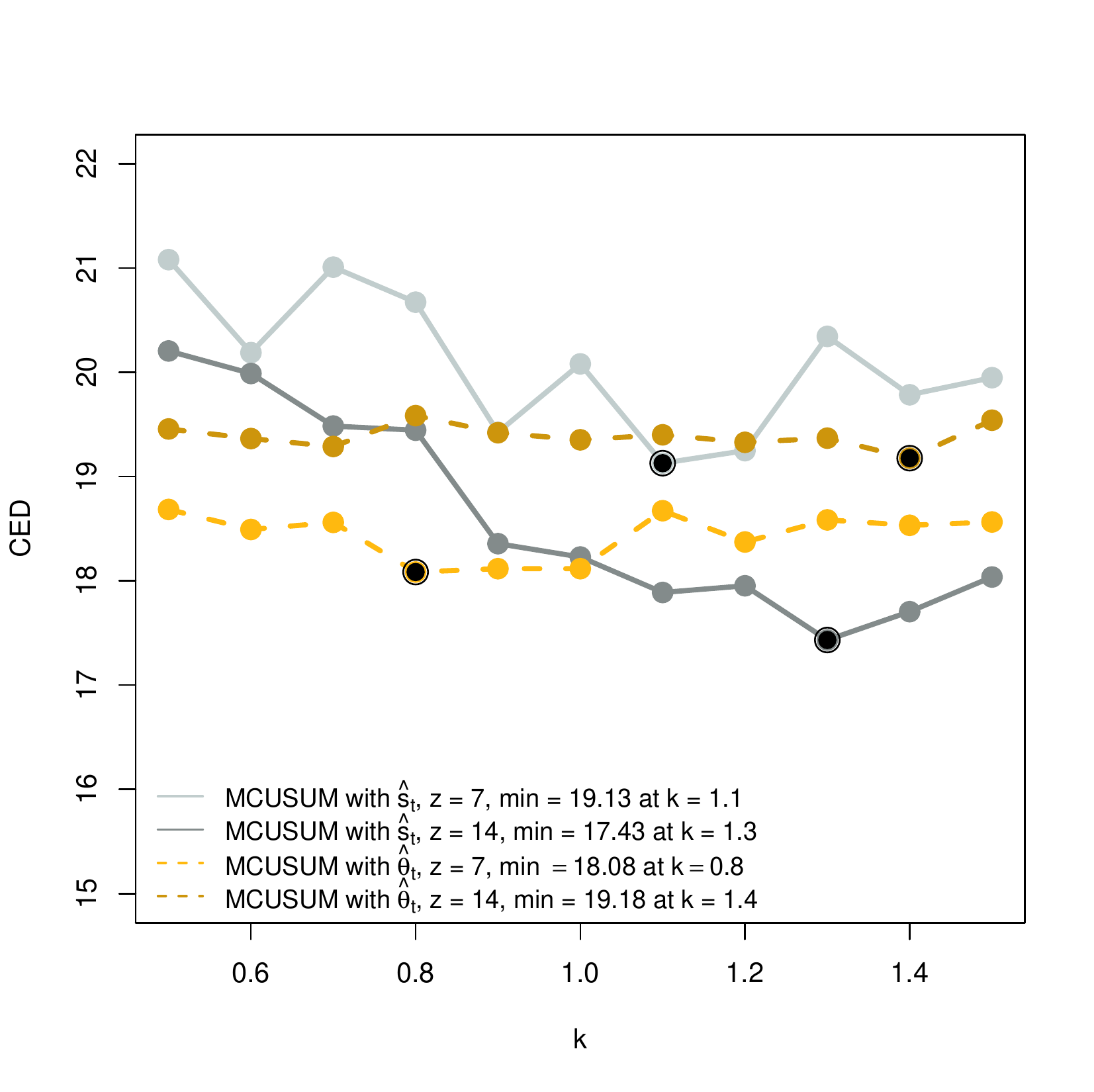}}
	\hfill
	\subfloat[MEWMA, Case B.1]{\includegraphics[width=0.5\textwidth, trim = 0cm 0.5cm 0cm 2cm, clip]{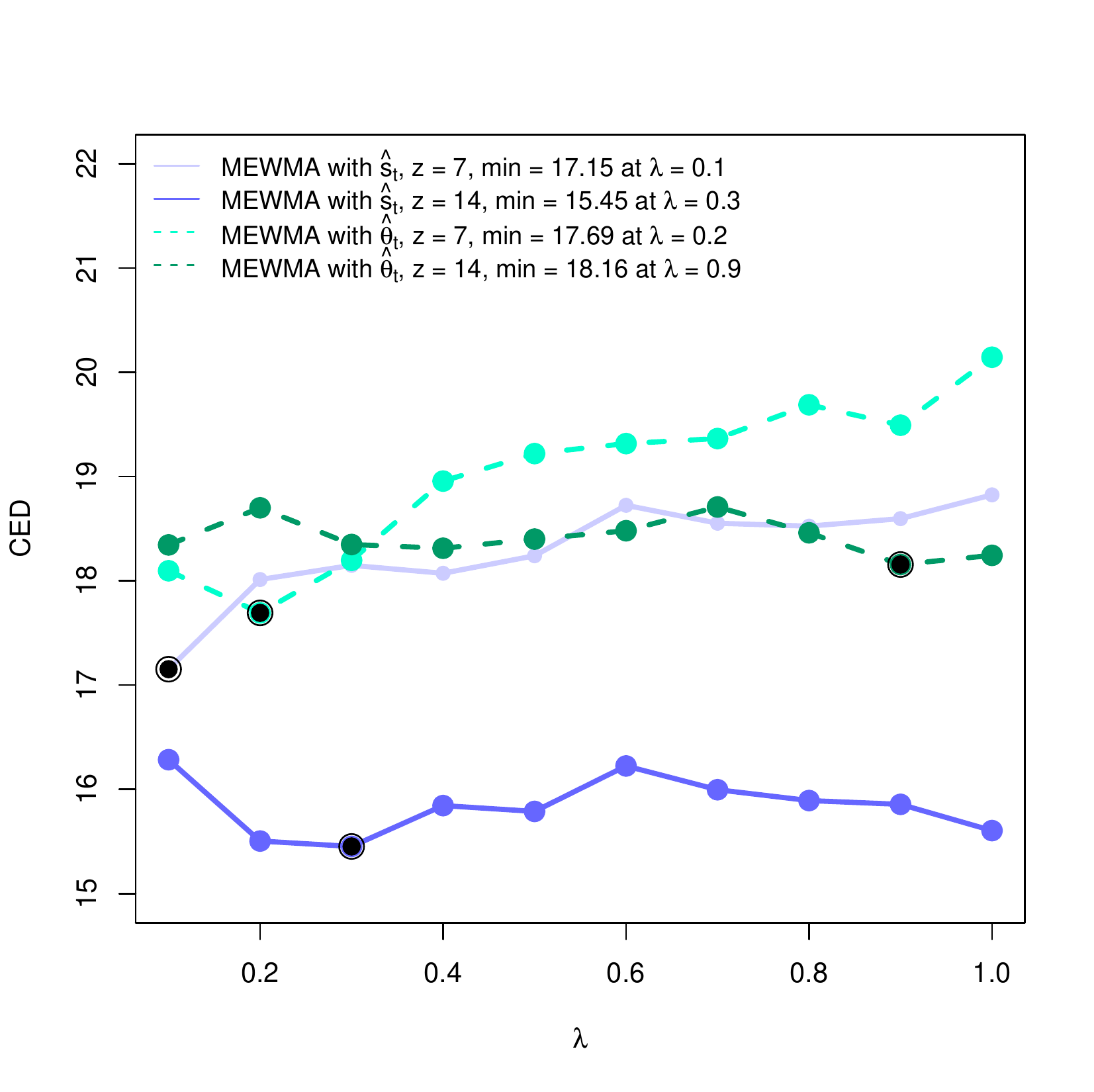}}
	\hfill
	\subfloat[MCUSUM, Case B.2]{\includegraphics[width=0.5\textwidth, trim = 0cm 0.5cm 0cm 2cm, clip]{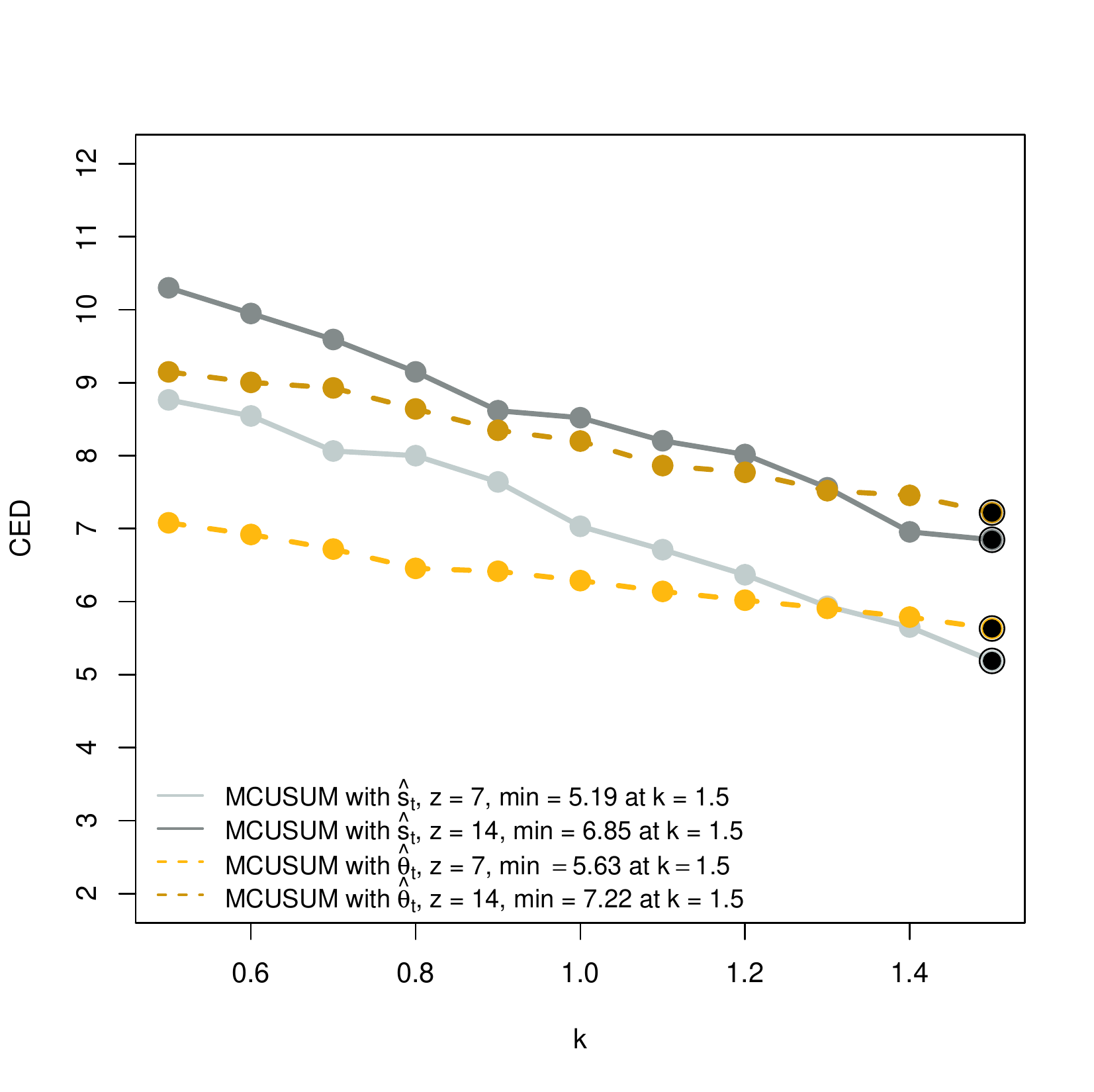}}
	\hfill
	\subfloat[MEWMA, Case B.2]{\includegraphics[width=0.5\textwidth, trim = 0cm 0.5cm 0cm 2cm, clip]{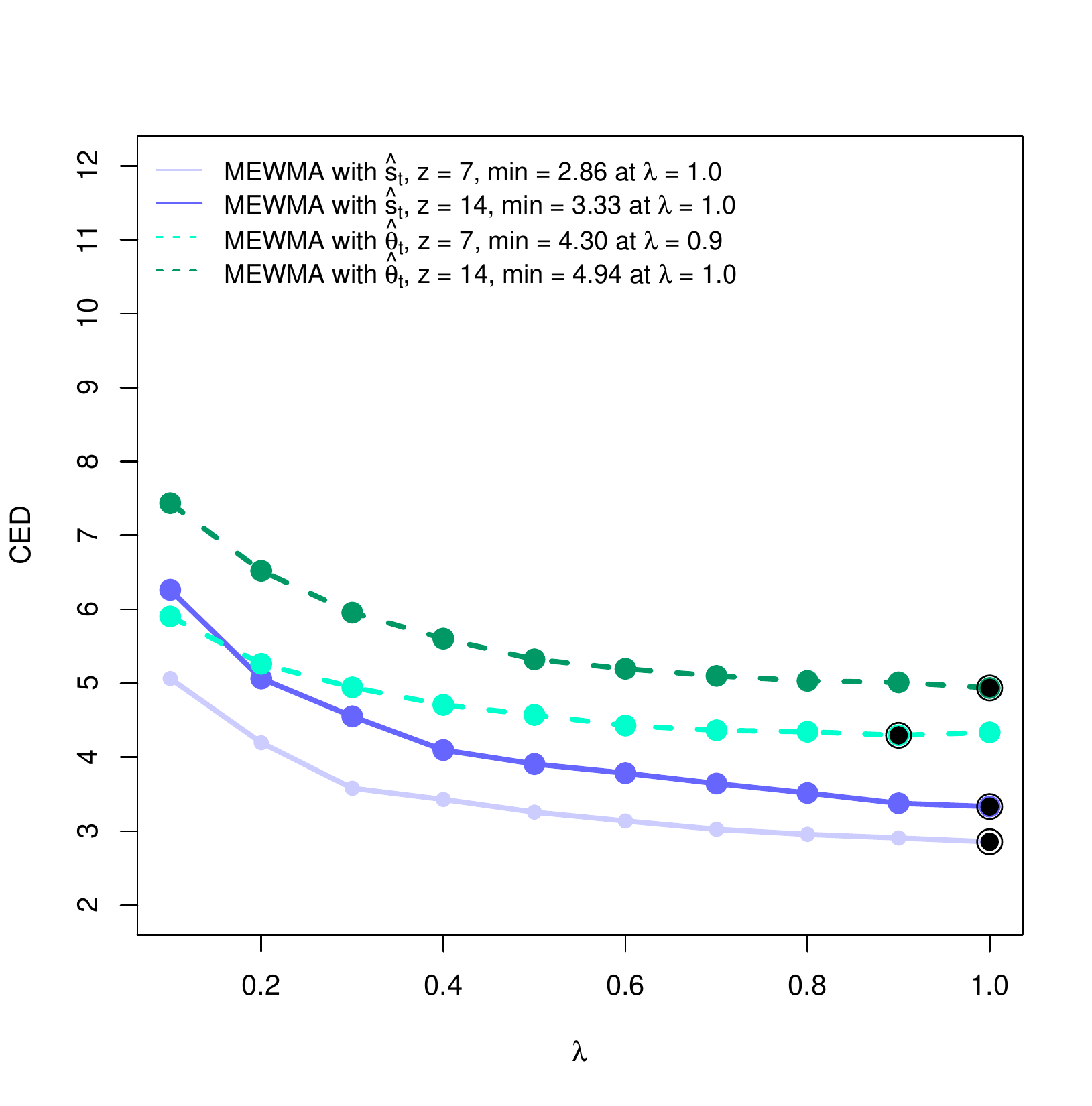}}
	\hfill
	\subfloat[MCUSUM, Case B.3]{\includegraphics[width=0.5\textwidth, trim = 0cm 0.5cm 0cm 2cm, clip]{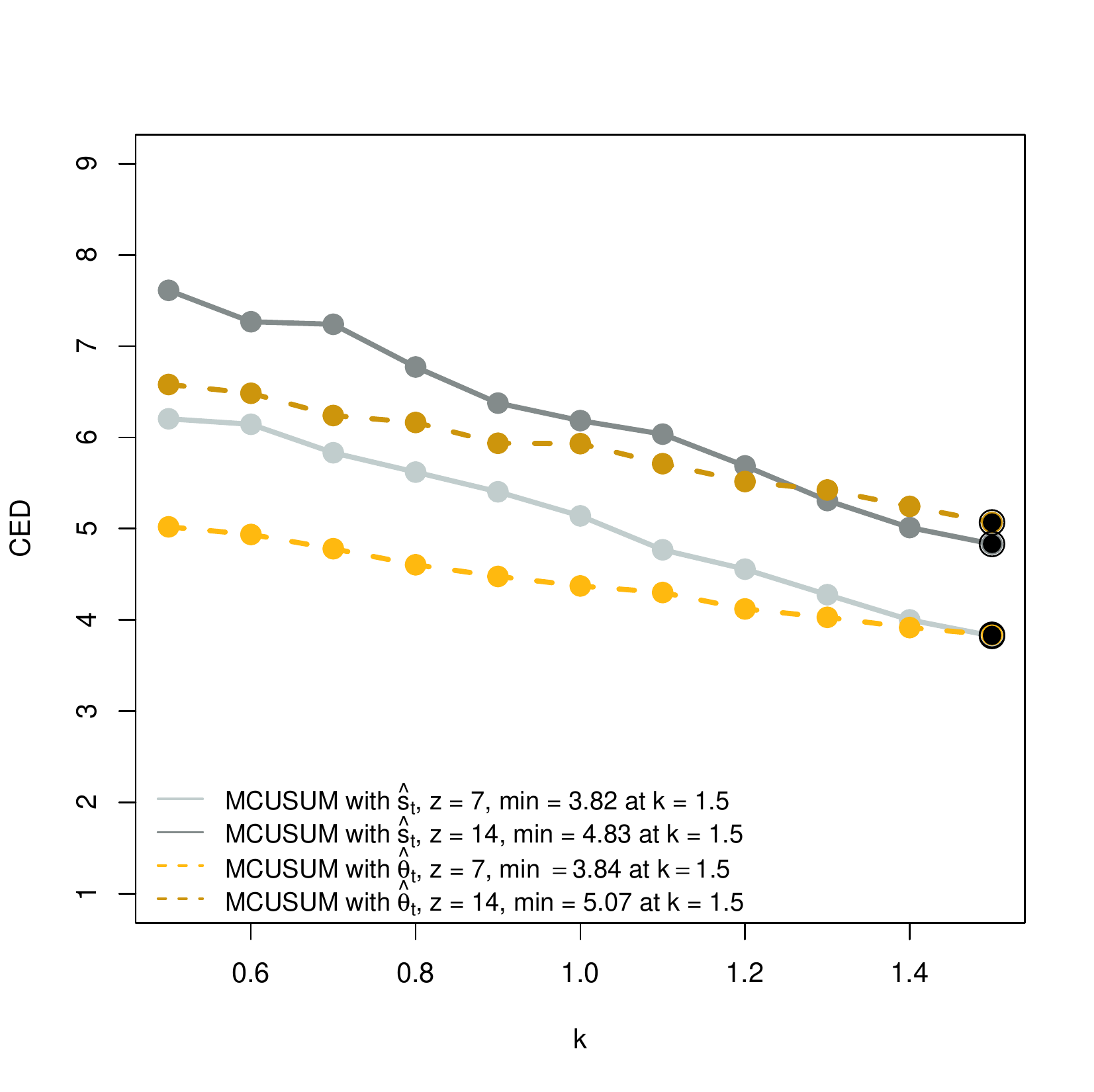}}
	\hfill
	\subfloat[MEWMA, Case B.3]{\includegraphics[width=0.5\textwidth, trim = 0cm 0.5cm 0cm 2cm, clip]{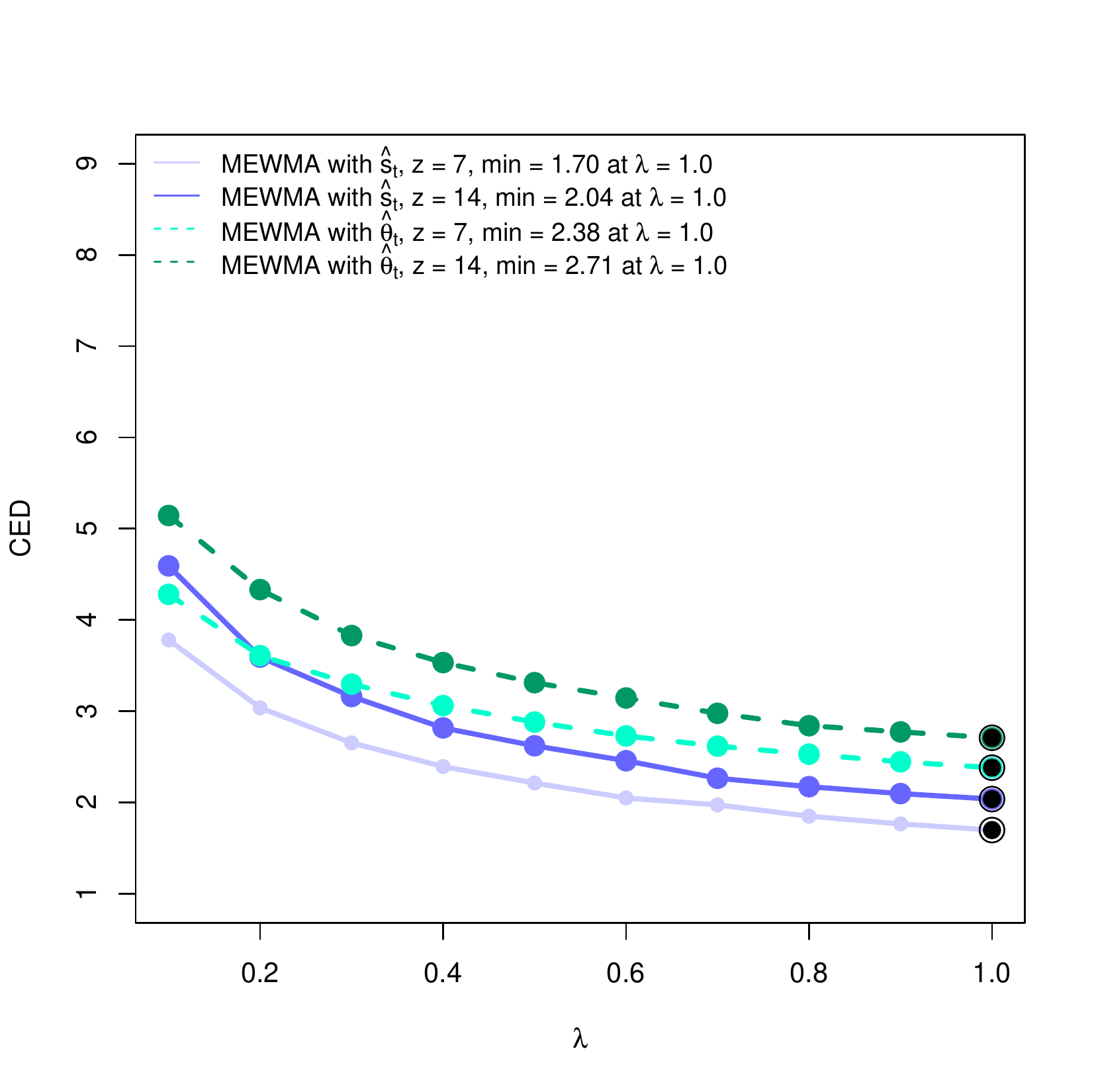}}
	
	\caption{{\color{blue}Conditional expected delays for anomalies of Type B for MCUSUM (left) and MEWMA (right) together with the different choices of the reference parameter $k$ and the smoothing parameter $\lambda$, the window sizes $z = 7$ and $z = 14$, and the network estimates $\hat{\bm{s}}_t$ (solid lines) and $\hat{\bm{\theta}}_t$ (dashed lines). Black points indicate the minimum CED for each setting.}}
	\label{PhiType2}
\end{figure}

\begin{figure}[H]
	\centering
	\subfloat[MCUSUM, Case C.1]{\includegraphics[width=0.5\textwidth, trim = 0cm 0.5cm 0cm 2cm, clip]{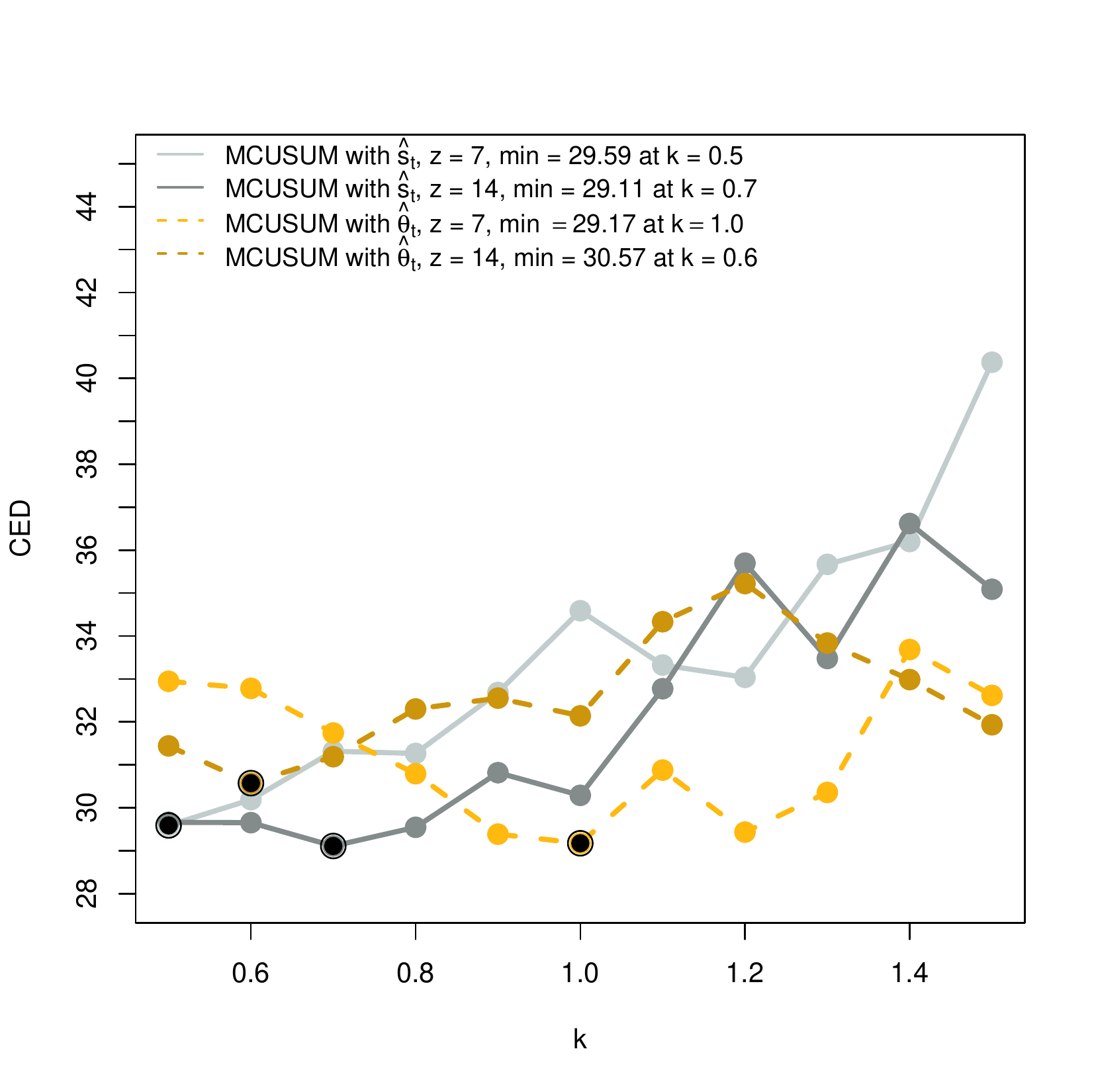}}
	\hfill
	\subfloat[MEWMA, Case C.1]{\includegraphics[width=0.5\textwidth, trim = 0cm 0.5cm 0cm 2cm, clip]{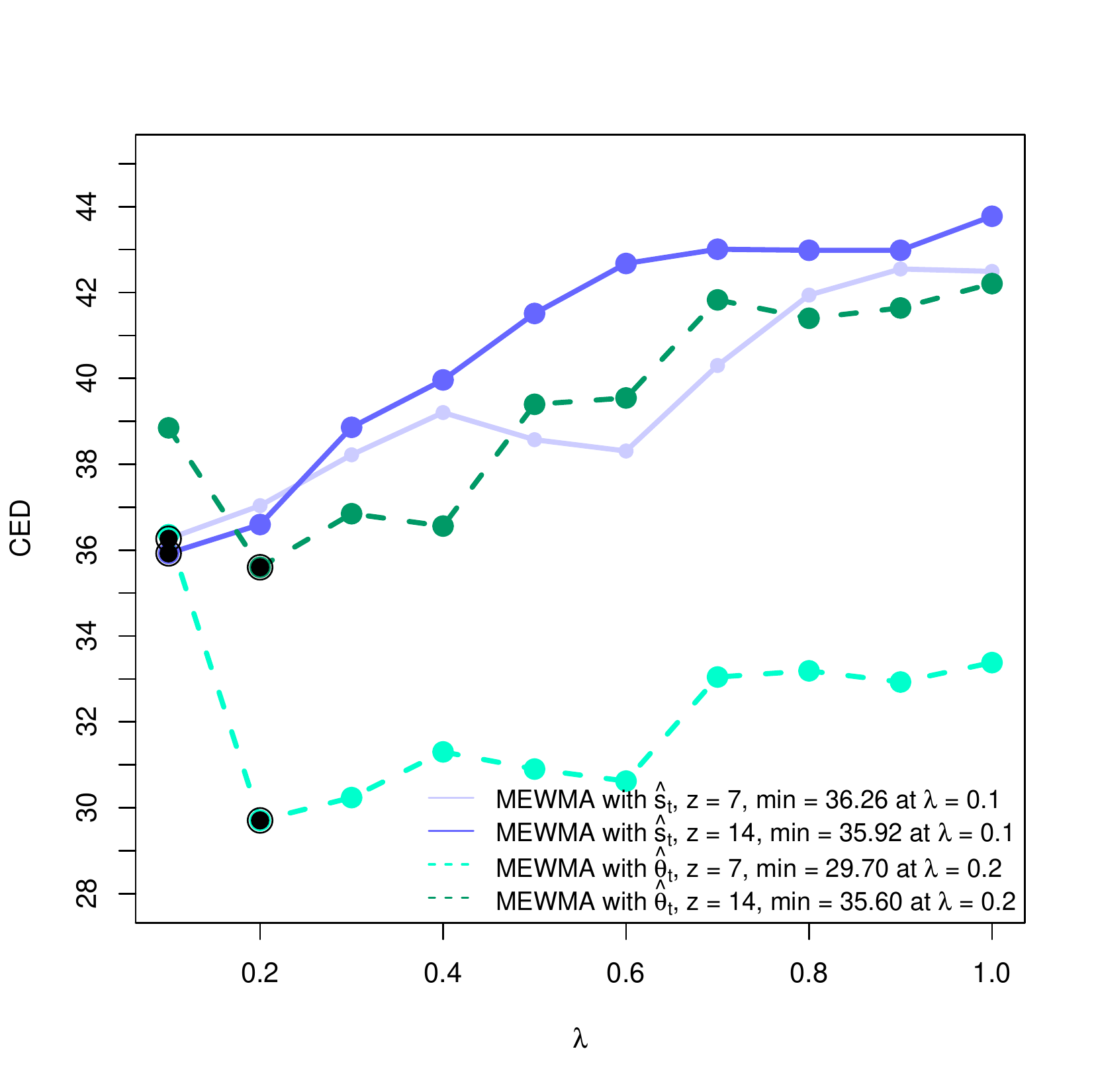}}
	\hfill
	\subfloat[MCUSUM, Case C.2]{\includegraphics[width=0.5\textwidth, trim = 0cm 0.5cm 0cm 2cm, clip]{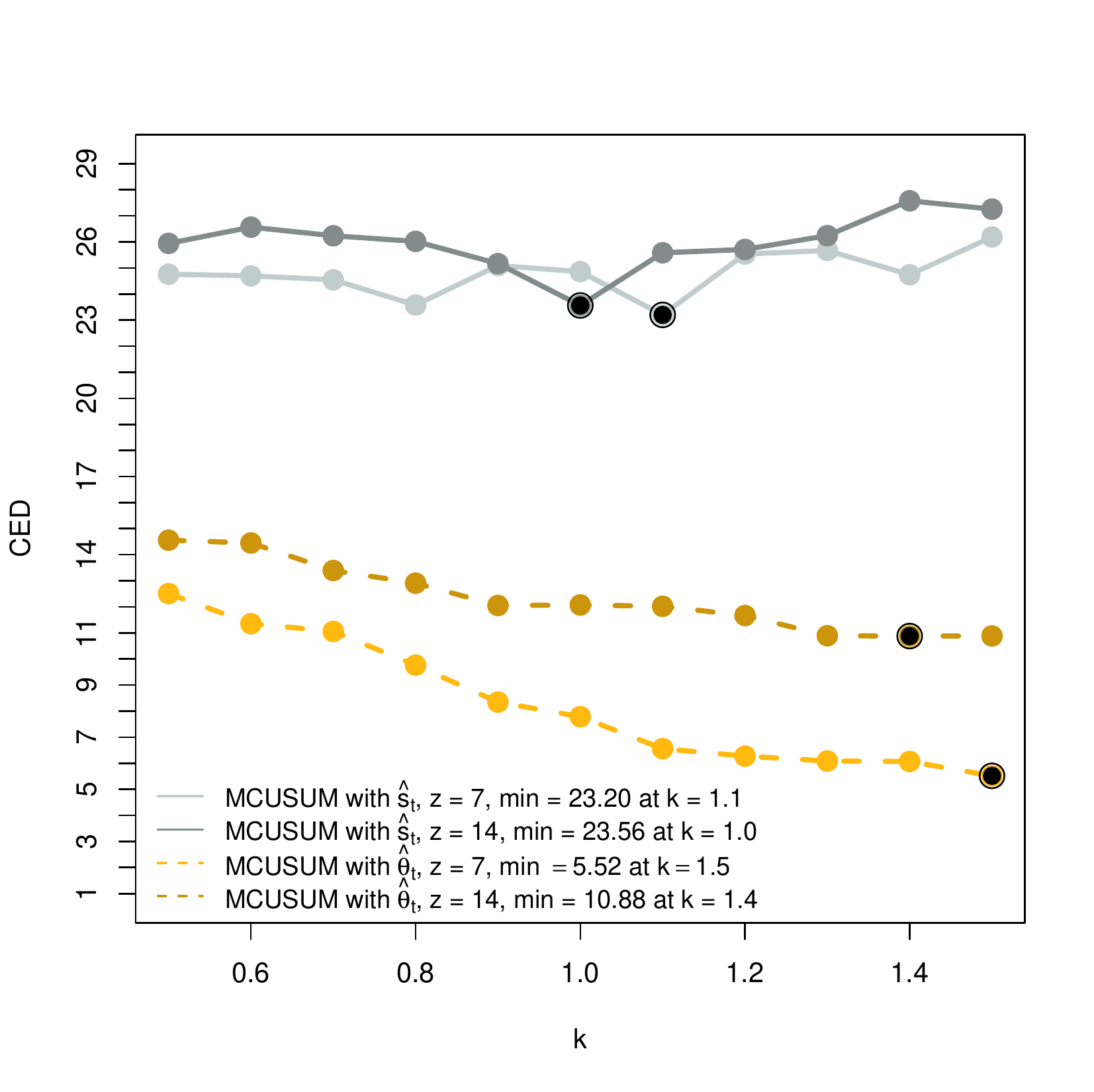}}
	\hfill
	\subfloat[MEWMA, Case C.2]{\includegraphics[width=0.5\textwidth, trim = 0cm 0.5cm 0cm 2cm, clip]{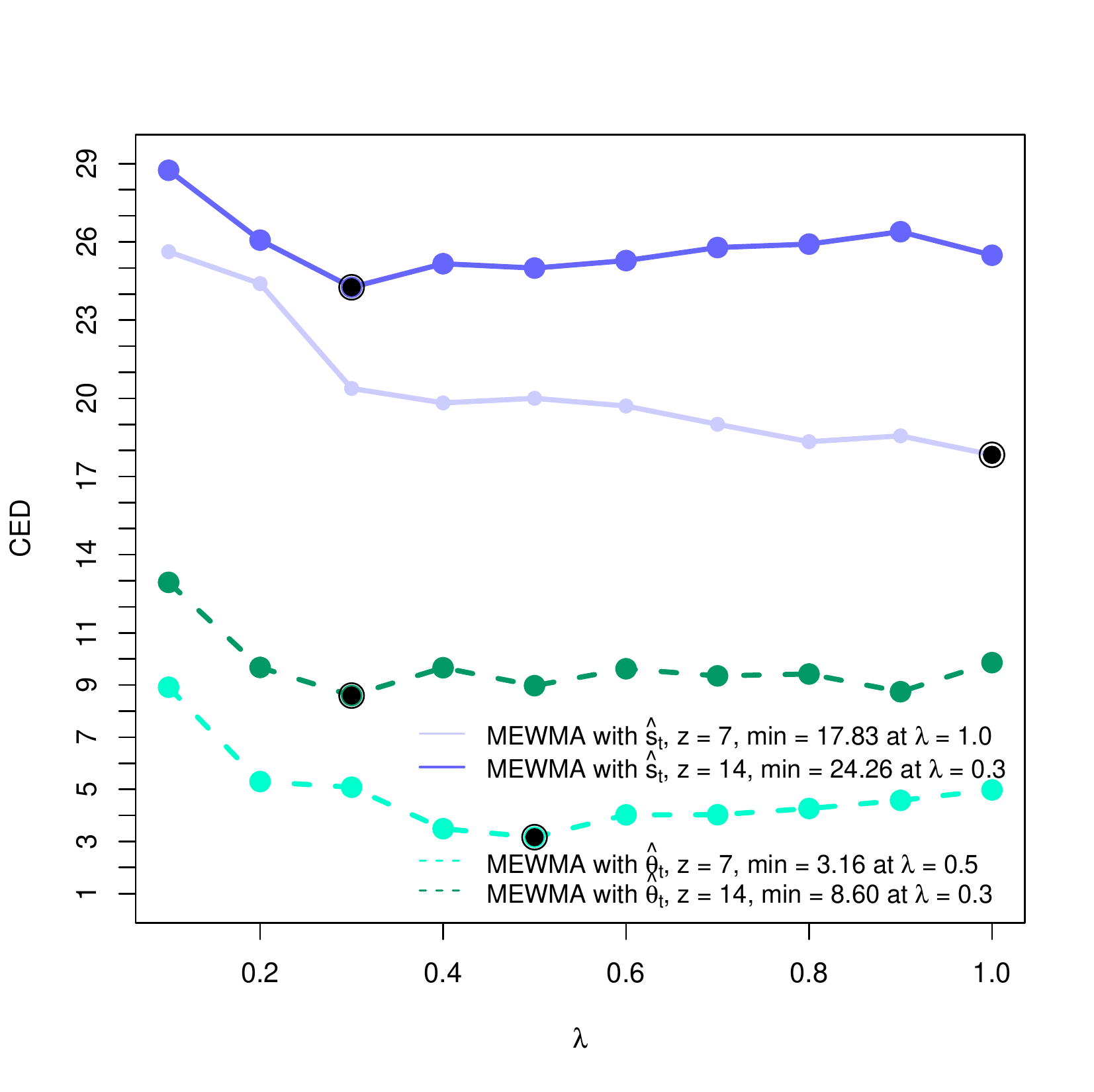}}
	\hfill
	\subfloat[MCUSUM, Case C.3]{\includegraphics[width=0.5\textwidth, trim = 0cm 0.5cm 0cm 2cm, clip]{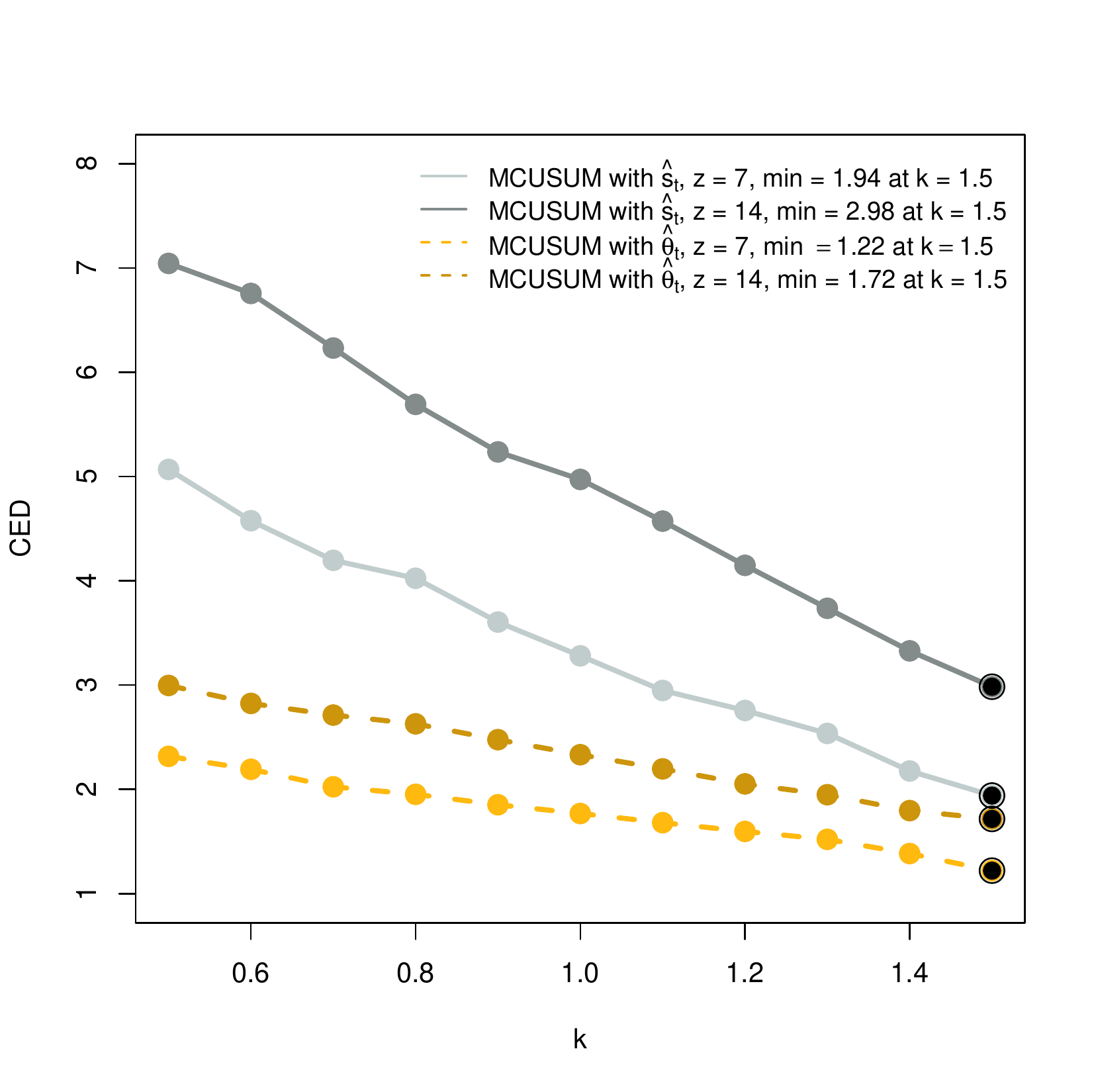}}
	\hfill
	\subfloat[MEWMA, Case C.3]{\includegraphics[width=0.5\textwidth, trim = 0cm 0.5cm 0cm 2cm, clip]{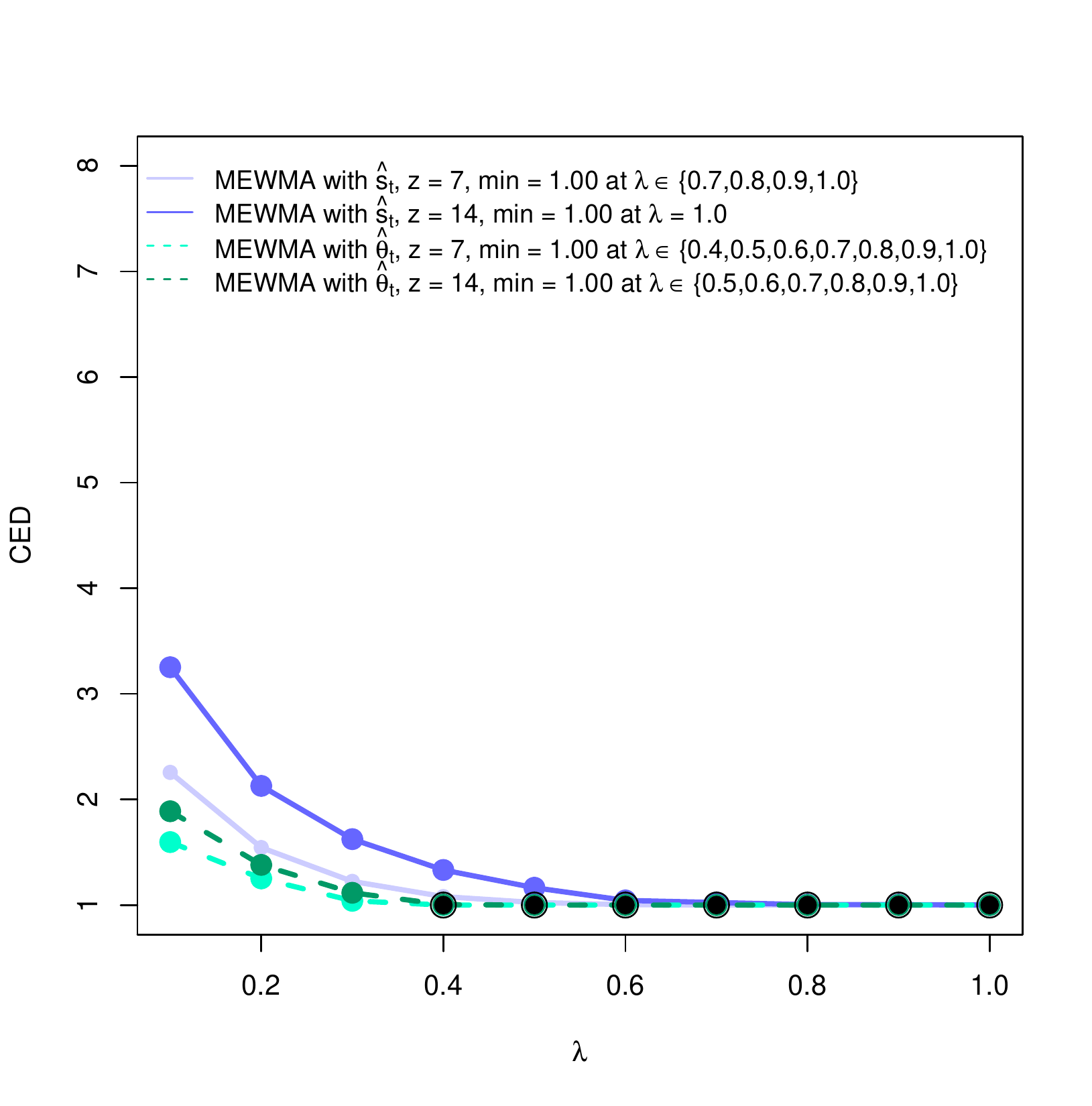}}
	
	\caption{{\color{blue}Conditional expected delays for anomalies of Type C for MCUSUM (left) and MEWMA (right) together with the different choices of the reference parameter $k$ and the smoothing parameter $\lambda$, the window sizes $z = 7$ and $z = 14$, and the network estimates $\hat{\bm{s}}_t$ (solid lines) and $\hat{\bm{\theta}}_t$ (dashed lines). Black points indicate the minimum CED for each setting.}}
	\label{MutualType3}
\end{figure}

In the next step, we analyse the performance of the proposed charts in terms of their detection speed. 
As a performance measure, we compute the conditional expected delay (CED) of detection, conditional on a false signal not having been occurred before the (unknown) time of change $\tau$ \citep{kenett2012assessing}. For our simulation, we set $\tau = 101$. Using 250 simulations, we estimate the CED based on the UCLs with $ARL_0 = 50$ for each setting. That means we would expect $\text{CED} = 50$ if no change happened and it should be considerably smaller in the case of an anomaly. Figures \ref{MType1}, \ref{PhiType2}  and \ref{MutualType3} present the results of the simulation for anomalies of Type A, B and C, respectively.

{\color{blue} There are several aspects to assess fully the obtained results. First of all, the comparison of performance between the MCUSUM and the MEWMA control charts.}
In most of the cases, the CED of the MEWMA chart is smaller compared to the corresponding MCUSUM chart. However, for the best choice of the reference parameter $k$ or the smoothing parameter $\lambda$, both charts are competitive. The respective values are indicated by the large dots indicating the minimum on the CED curve. {\color{blue} For instance, the weakest change of Type A.1 (Figure \ref{MType1}, (a)) is detected quicker by the MCUSUM chart with the low parameters $k$.}  In contrast, the MEWMA charts perform better for {\color{blue} bigger changes such as in Cases 2 and 3.} 

Generally speaking, we see that the CED is decreasing if the shift size or the intensity of the change is increasing. Moreover, if the reference parameter $k$ or the smoothing parameter $\lambda$ is smaller, less intense anomalies can be detected. If {\color{blue}in practical implementation the detection of larger changes is required}, these parameters should also be higher. It is worth reminding that the MEWMA charts coincide with Hotelling's chart if $\lambda = 1$, i.e., the control statistic depends only on the current value.

The disadvantage of both approaches is that small and persistent changes are not detected quickly when the parameters $k$ or $\lambda$ are not optimally chosen. For example, considering Case A.2 in Figure \ref{MType1} (b), we can notice that at the high values of the parameter $\lambda$ the CED slightly exceeds the $ARL_0$ reflecting the poor performance. However, a careful selection of the parameters can overcome this problem.  {\color{blue}Also, the choice of the window size plays a significant role in detecting the anomalies reliably, being a trade-off between a precise description of the process and the ability to reflect the sudden changes in its behaviour.} 

{\color{blue}Regarding the differences in results with respect to the quantities $\hat{\bm{\theta}}_t$ and  $\hat{\bm{s}}_t$, we notice a similar performance in Anomaly Types A and B. Interestingly in most of the cases the MEWMA control charts work better for $\hat{\bm{s}}_t$ and the CUSUM control charts for $\hat{\bm{\theta}}_t$. However, looking at the detection of anomaly Type C.2, we observe a considerable advantage of applying $\hat{\bm{\theta}}_t$ rather than $\hat{\bm{s}}_t$. To confirm that this behaviour is supported by another example, we created an additional test case with $\zeta = 0.02$. These results as well as the others from Type C anomalies are summarised in Table \ref{CED_Type3}. As we can observe, if the change is too small, then both groups of control charts created on the basis of $\hat{\bm{\theta}}_t$ and $\hat{\bm{s}}_t$ need relatively long to detect it. In case when $\zeta = 0.05$, representing a substantial anomaly, the change is identified quickly independent from the choice of the estimates. However, when the change is of a moderate degree, for example, $\zeta = 0.02$, then the control charts based on $\hat{\bm{\theta}}_t$ signal the anomalous behaviour considerably quicker. Whether the main reason for such difference is the particular type of anomaly, namely it is an example of a point change, cannot be said generally as additional variations of such anomalies should be  examined. However, from the evidence in this work, the authors hypothesise that the estimates $\hat{\bm{\theta}}_t$  might be more suitable for general network monitoring when it is assumed that a point, as well as a persistent change can occur, though the comparison between the performance of $\hat{\bm{\theta}}_t$ and $\hat{\bm{s}}_t$ is worth further investigation.}

\begin{table}[H]
	\centering
	\small
	\begin{tabular}{p{4cm}lcccclcccc}
		
		\toprule
		Estimates    &  CED  &      &    &\hspace{-1cm} $\hat{\bm{s}}_t$ &    &  &  & & \hspace{-1cm} $\hat{\bm{\theta}}_t$ \tabularnewline
		\midrule
		Case    &   &  C.1   &   C.2    &   &   C.3 &   &  C.1     &   C.2    &  &   C.3 \tabularnewline
		Parameter $\zeta$    &   &  \textbf{0.005}   &   \textbf{0.01}    &   \textbf{0.02}    &   \textbf{0.05} &   &  \textbf{0.005}    &   \textbf{0.01}    &   \textbf{0.02}    &   \textbf{0.05} \tabularnewline [0.2cm]
		
		MEWMA with $z = 7$         & Min. & 36.26&17.83&  1.85    &   \emph{1.00}    &    &  \emph{29.70}&  \emph{3.16} & \emph{1.00} &\emph{1.00}\tabularnewline
		& $\lambda_{min}$ & 0.1    &  1.0 &  1.0    &   0.7   & &  0.2  & 0.5& 0.9& 0.4\\
		&Max. & 42.55    &    25.62  &  7.16  &   2.26   &    &  36.36&  8.92 & 2.56 &1.60\tabularnewline
		& $\lambda_{max}$ & 0.9   &    0.1  &  0.1    &  0.1   &    &  0.1&  0.1 & 0.1 &0.1\\[.1cm]
		MEWMA with $z = 14$         & Min. & 35.92    &   24.26  &  3.57    &   \emph{1.00}    &    &  35.60&  8.60 & \emph{1.00} &\emph{1.00}\tabularnewline
		& $\lambda_{min}$ & 0.1    &    0.3  &  0.9    &   1.0    &    &  0.2&  0.3 & 1.0 &0.5\\
		
		&Max. & 43.78    &   28.75  &  9.90   &   3.25    &    &  42.21&  12.94 & 3.49 &1.89\tabularnewline
		& $\lambda_{max}$ & 1.0   &    0.1  &  0.1    &   0.1    &    &  1.0& 0.1 & 0.1 &0.1\\[.1cm]

		\midrule
		
		MCUSUM with $z = 7$&  Min.   &   29.59  &  23.20  &   6.66   &  1.94    &  & 29.17 &\emph{5.52}&\emph{2.10}&\emph{1.22} \tabularnewline
		& $k_{min}$ & 0.5    &    1.1  &  1.5    &   1.5    &    &  1.0&  1.5 & 1.5 &1.5\\
		&  Max.   &    40.38   &  26.19    &   18.70     &  5.07    &  &  33.68&  12.51&3.91&2.32 \tabularnewline
		& $k_{max}$ & 1.5   &   1.5  &  0.5    &   0.5    &    & 1.4&  0.5 & 0.5 &0.5\\[.1cm]
		MCUSUM with $z = 14$ &  Min.    &   \emph{29.11}    &  23.56     &  9.45     &    2.98  &  &30.57  &10.88&3.14&1.72    \tabularnewline  
		& $k_{min}$ & 0.7    &   1.0  & 1.5   &   1.5    &    &  0.6&  1.4 & 1.5 &1.5\\
		&  Max.    &   36.62  &  27.58    &   18.86     & 7.04     &  & 35.22 &  14.56&6.19&3.00
		\tabularnewline
		& $k_{max}$ & 1.4    &    1.4 &  0.5    &   0.5    &    &  1.2&  0.5 & 0.5 &0.5\\
		
		\bottomrule
	\end{tabular}
	\caption{{\color{blue}Summary of the CED results to detect anomalies of Type C with the additional test case $\zeta = 0.02$. The corresponding smoothing and reference parameters $\lambda$ and $k$ are provided under the respective CED. The minimum CED for each case and the control chart group are underlined. The maximum CED represents the ``worst-case'' scenario. In case several values of the parameter $\lambda$ correspond to the CED result, only the smallest value is reported.}}
	\label{CED_Type3}
\end{table}

To summarise, the effectiveness of the presented charts to detect structural changes depends significantly on the accurate estimation of the anomaly size one aims to detect.
Thus, to ensure that no anomalies were missed, it can be effective to apply paired charts and benefit from the strengths of each of them to detect varying types and sizes of anomalies, if the information on the possible change is not available or not reliable.

\begin{figure}[H]
	\begin{center}
	\mbox{} \hspace{0.8cm} April 1, 2018 \hspace{4cm} April 1, 2019 \hspace{3cm} April 1, 2020 \hfill \mbox{}\\
		
		\includegraphics[trim= 40 100 0 40,clip, width=1.0\textwidth]{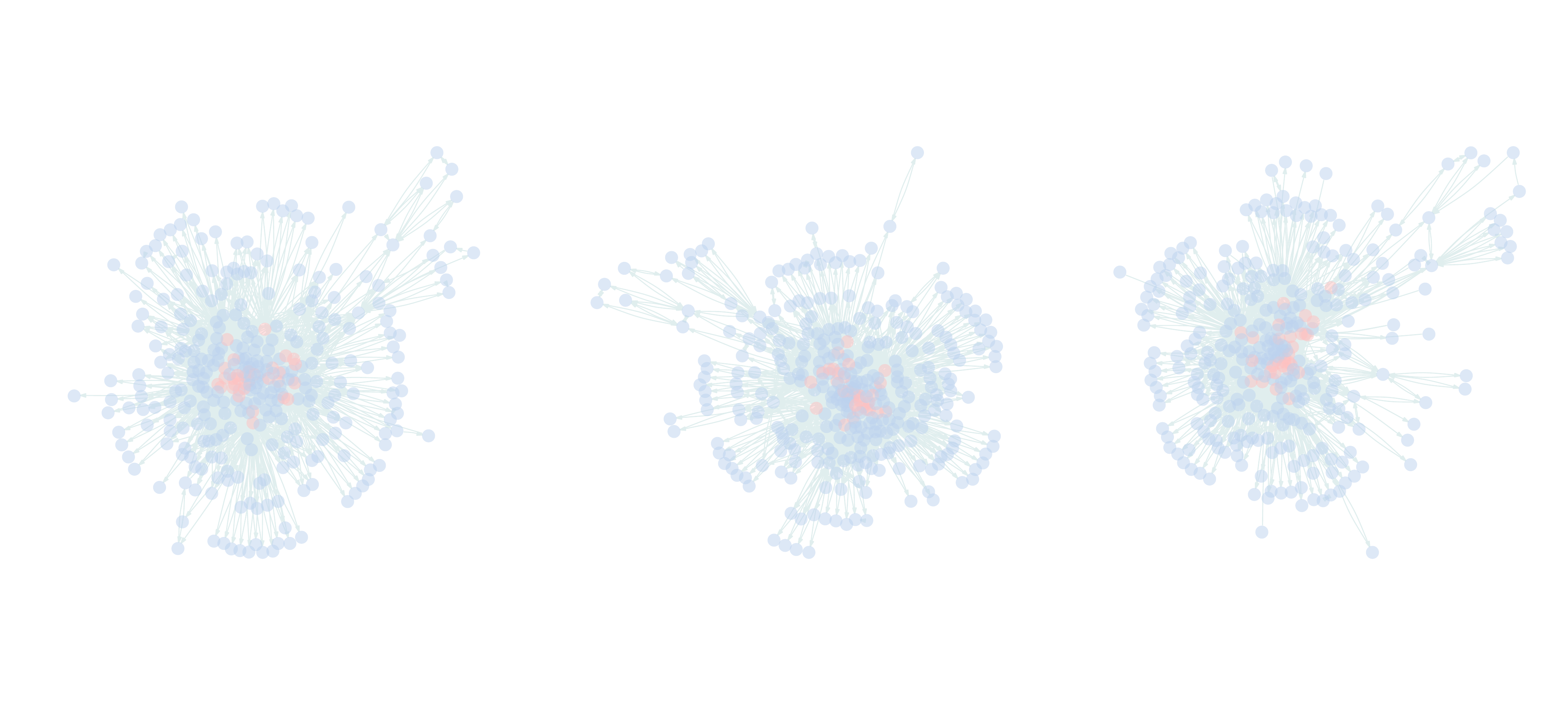}\\[1cm]
		
		\caption{Illustration of the flight network on April 1 of each year excluding isolated vertices. It can be seen that the topology of the network has changed. The red coloured nodes represent the 30 busiest airports.}\label{fig:net_flights}	
	\end{center}
\end{figure}

\section{Empirical Illustration} \label{section:4}

To demonstrate applicability of the described method, {\color{blue} we monitor} the daily flight data of the United States (US) published by the US Bureau of Transportation Statistics {\color{blue} using the parameter estimates $\hat{\bm{\theta}}_t$}. Each day can be represented as a directed network, where nodes are airports and directed edges define flights between airports. In Figure \ref{fig:net_flights}, examples of flight network data in 2018, 2019 and 2020 (until the end of April) are presented.

Flexibility in choosing the {\color{blue} network terms}, according to the type of anomalies one would like to detect, enables different perspectives on the same network data. In our case, we aim to identify considerable changes the network development. The intuition of how the flight network usually operates guides the choice of its terms. At the time of writing, due to the current COVID-19 pandemic in the year 2020, some regions have paused the operation of transport systems with the aim to reduce the number of new infections. However, the providers enable mobility by establishing connections through territories which allow travelling. That means, instead of having a direct journey from one geographical point to another, currently the route passes through several locations, which can be interpreted as nodes. Thus, the topology of the graph has changed: instead of directed mutual links, the number of intransitive triads and asymmetric links starts to increase significantly. We can incorporate both terms, together with the edge term and a memory term ($v = 1$), and expect the estimates of the respective coefficients belonging to the first two statistics to be close to zero or strongly negative in the in-control case.

\begin{figure}[H]
	\begin{center}
		\includegraphics[width=0.5\textwidth, trim= 0 0 10 0,clip]{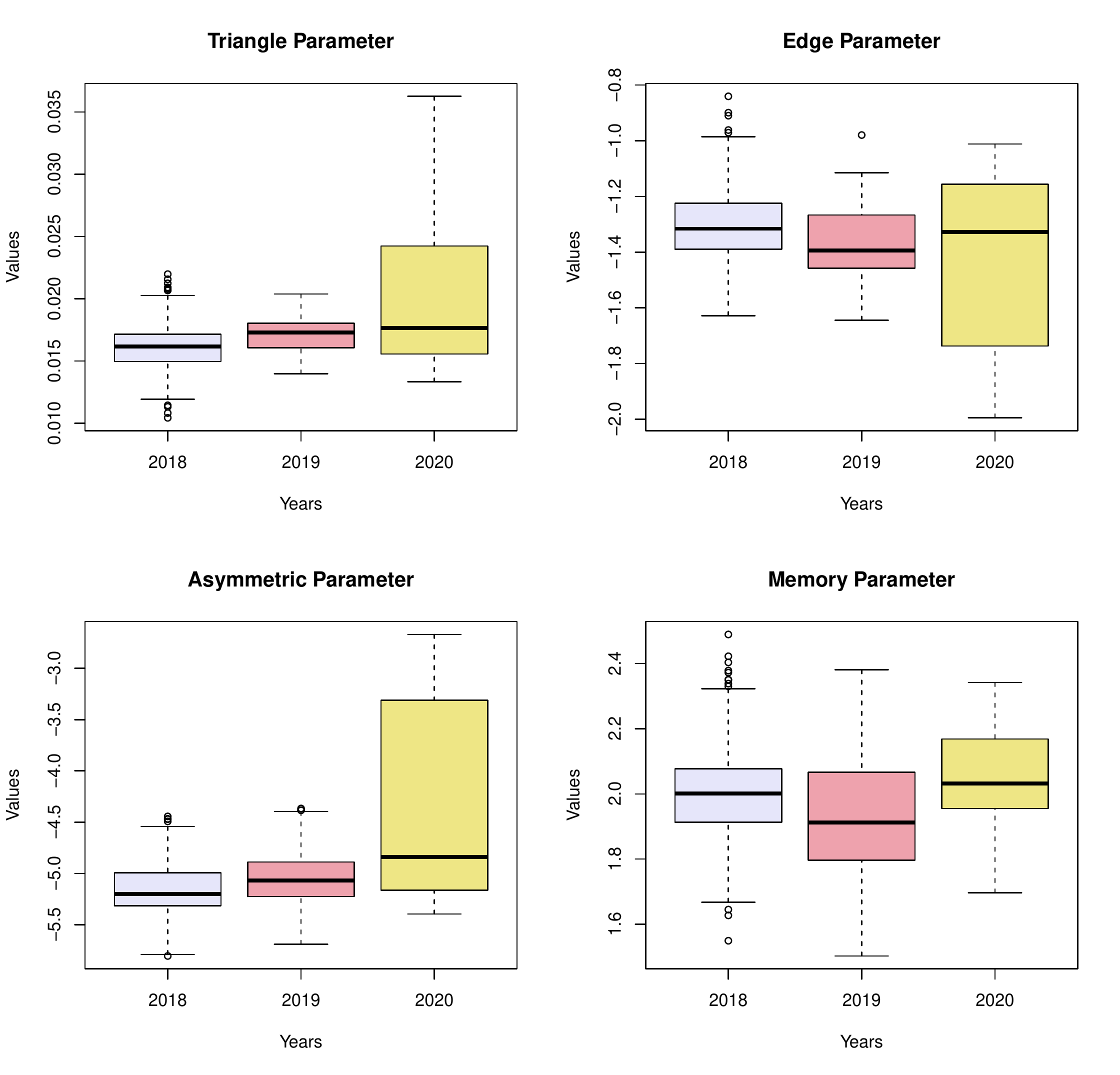}
		\caption{Distribution of the estimated coefficients $\hat{\bm{\theta}}_t$ in 2018, 2019 and 2020.}
		\label{Boxplots}
	\end{center}
\end{figure}

Initially, we need to decide which data are suitable to define observations coming from Phase I, i.e., the in-control state. There were no considerable events which would seriously affect the US flight network known to the authors in the year 2018, therefore, we chose this year to characterise the in-control state. Consequently, the years 2019 and 2020 represent Phase II. To capture the weekly patterns, a time window of size $z = 7$ was chosen, so that the first instant of time represents January 8, 2018. In this case, Phase I consists of 358 observations and Phase II of 486 observations. To guarantee that only factual flight data are considered, we remove cases when a flight was cancelled. {\color{blue} Additionally, we eliminate multiple edges.} The main descriptive statistics for Phase I and II are reported  in Table \ref{Stats}. There are no obvious changes when considering the descriptive statistics. Hence, control charts, which are only based on such characteristics, could fail to detect the possible changes in 2019 and 2020. When considering the estimates $\hat{\bm{\theta}}_t$ of the TERGM described by a series of boxplots in Figure \ref{Boxplots}, we can observe substantial changes in the values.

\begin{table}
	\centering
	\small
	\begin{tabular}{p{3cm}lccc}
		
		\toprule
		&          &   2018     &   2019    &   2020     \\
		\midrule
		Phase             &          &   I & II & II  \\[.1cm]
		Number of nodes   &          &    358     &   360     &    354     \\[.1cm]
		&  Min.    &    0.031   &  0.033    &   0.022    \\
		Density           &  Median  &    0.037   &  0.038    &   0.038    \\
		&  Max.    &    0.039   &  0.040    &   0.041    \\[.1cm]
		&  Min.    &    0.97    &  0.96     &   0.89     \\
		Reciprocity       &  Median  &    0.99    &  0.99     &   0.99     \\
		&  Max.    &    1.00    &  1.00     &   1.00     \\[.1cm]
		&  Min.    &    0.315   &  0.322    &   0.263    \\
		Transitivity      &  Median  &    0.339   &  0.339    &   0.326    \\
		&  Max.    &    0.357   &  0.354    &   0.345    \\
		\bottomrule
	\end{tabular}
	\caption{Descriptive statistics of the US flight network data. Density is calculated on networks without multiple edges.}
	\label{Stats}
\end{table}

\begin{figure}[H]
	\begin{center}
		\includegraphics[width=0.7\textwidth]{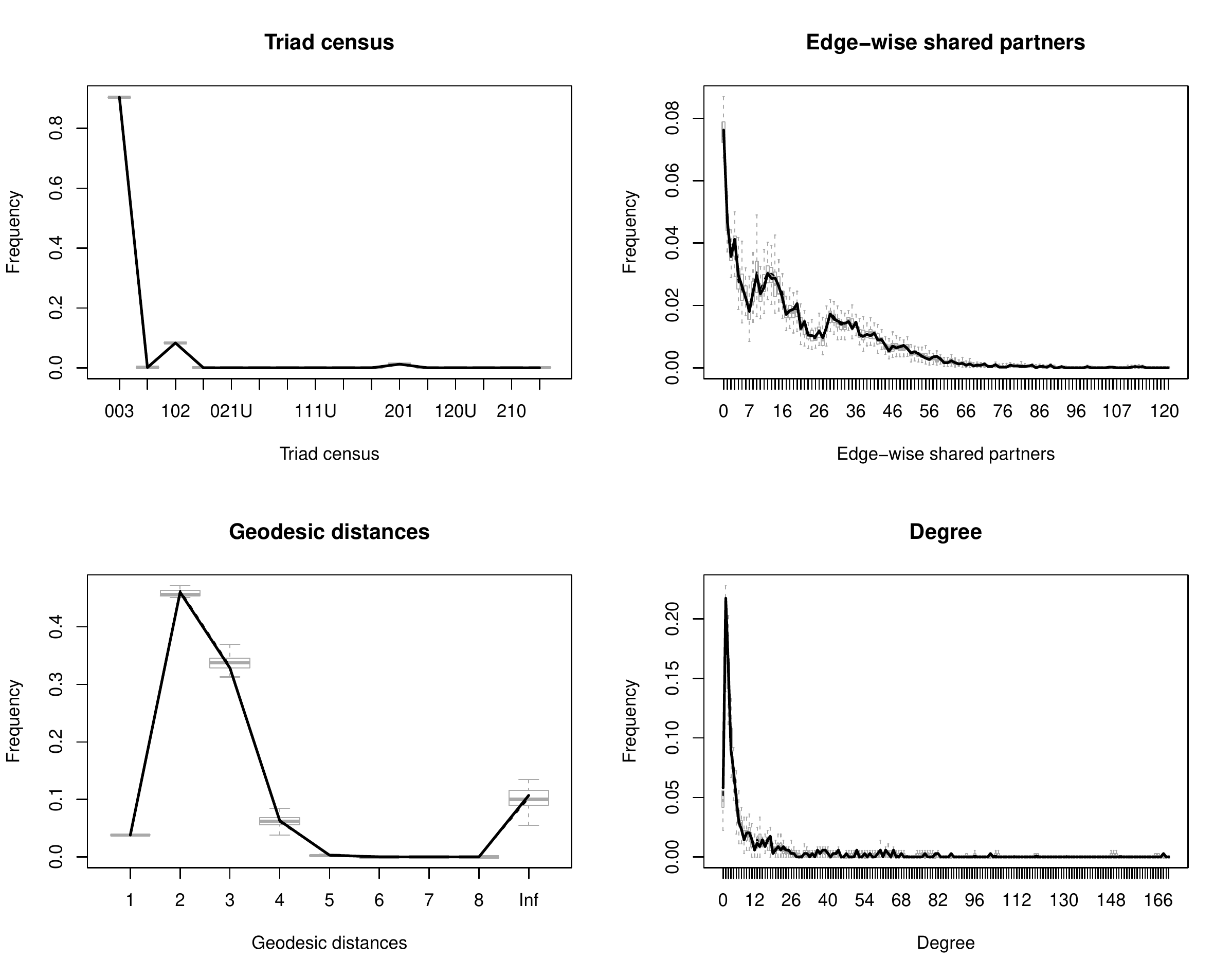}
		\caption{Illustration of the goodness-of-fit assessment for the TERGM. The considered networks belong to the period April 3-9 2019.}
		\label{fig:gof}
	\end{center}
\end{figure}

Before proceeding with the analysis, it is important to evaluate whether a TERGM fits the data well \citep{hunter2008goodness}. For each of the years, we randomly selected one period of the length $z$ and simulated 500 networks based on the parameter estimates from each of the corresponding networks. Figure \ref{fig:gof} depicts the results for the time frame April 3-9 2019, where the grey boxplots of each of the statistics represent the simulations, and the solid black lines connect the median values of the observed networks. Despite the relatively simple definition of the model, some typical network characteristics such as the distributions of edge-wise shared partners, the vertex degrees, various triadic configurations (triad census) and geodesic distances (the value of infinity replicates the existence of isolated nodes) match the observed distributions of the same statistics satisfactory.

To select appropriate control charts, we need to take into consideration the specifications of the flight network data. Firstly, it is common to have 3-4 travel peaks per year around holidays, which are not explicitly modelled, so that we can detect these changes as verifiable anomalous patterns. It is worth noting that one could account for such seasonality by including nodal or edge covariates. Secondly, as we aim to detect considerable deviations from the in-control state, we are more interested in sequences of signals. Thus, we have chosen $k = 1.5$ for MCUSUM and $\lambda = 0.9$ for the MEWMA chart. The target $ARL_0$ is set to 100 days, therefore, we can expect roughly 3.65 in-control signals per year by the construction of the charts.


\begin{figure}[H]
	\begin{center}
		\includegraphics[width=0.8\textwidth]{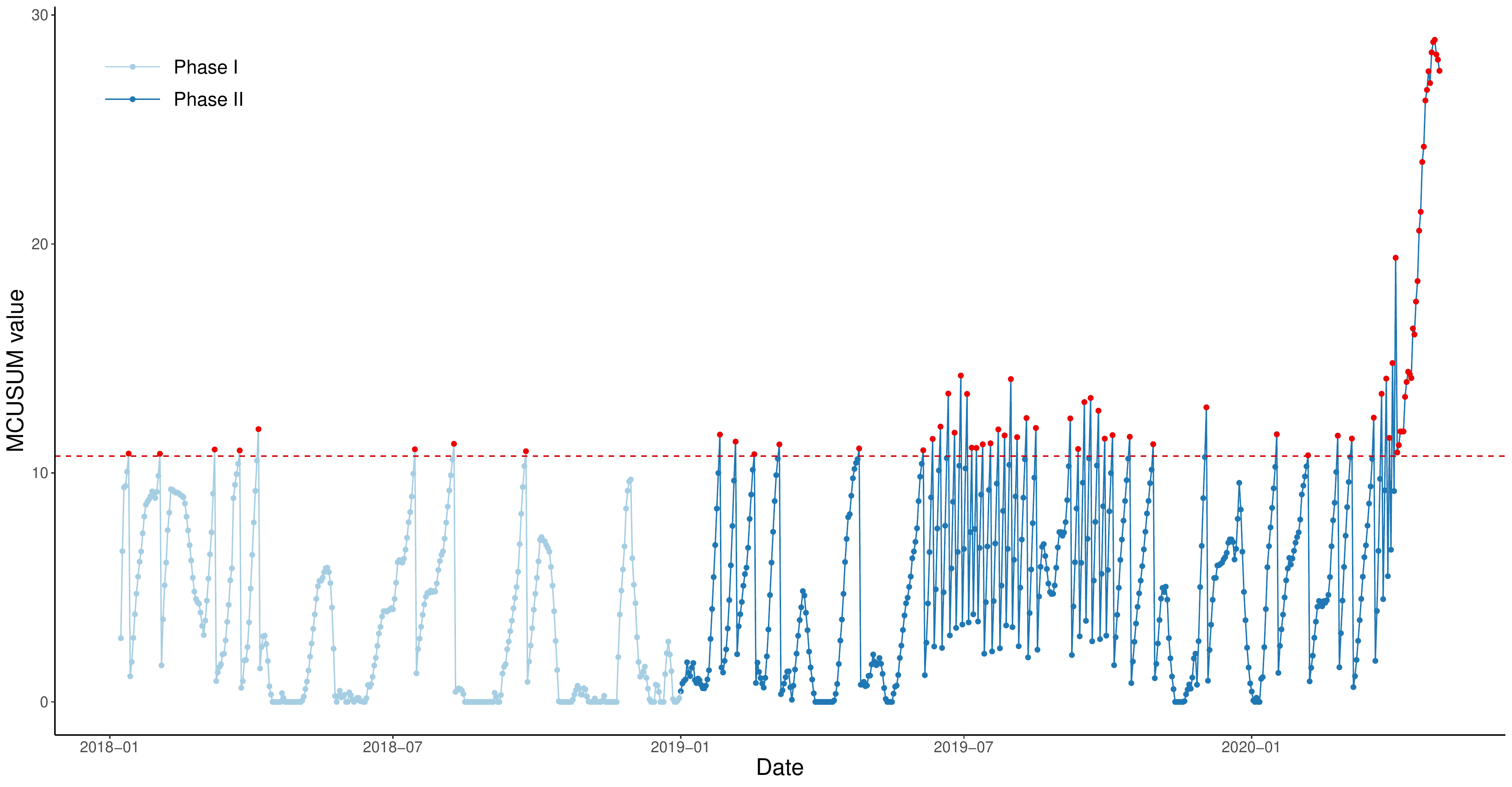}\\
		\includegraphics[width=0.8\textwidth]{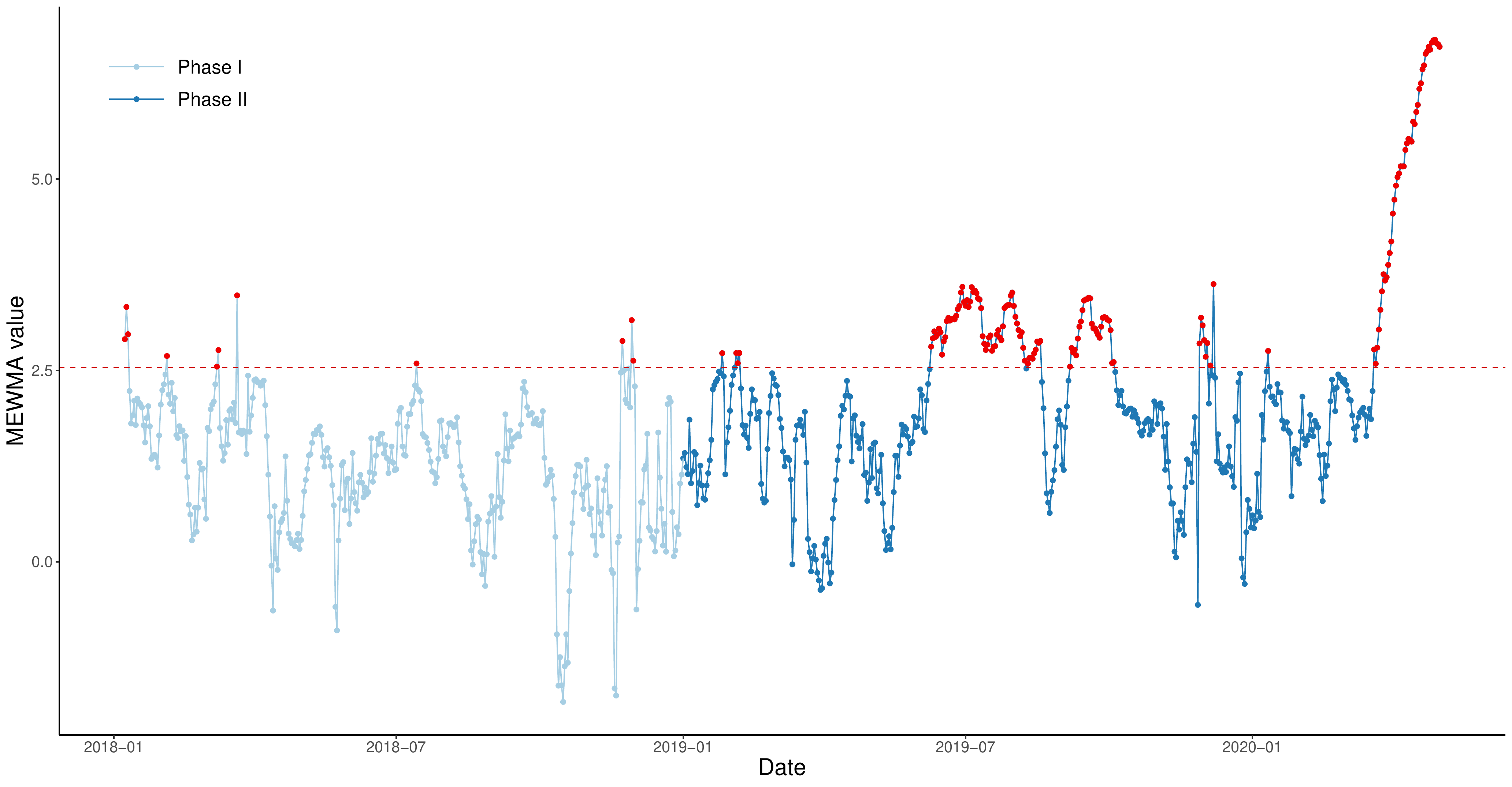}
		\caption{{\color{blue}The MCUSUM control chart (above) and the logarithmic MEWMA control chart (below). The horizontal red line corresponds to the upper control limit and the red points to the occurred signals.}}
		\label{fig:Charts}
	\end{center}
\end{figure}

Figure \ref{fig:Charts} depicts the results of both charts for monitoring the US flight network data. In Phase I there are slightly more in-control signals than expected, which we leave without investigation as they occur as singular instances. Considering Phase II, there are several anomalous behaviours which were detected. The first series of signals in summer 2019 is due to a particularly increased demand for flights during the holidays. The second sequence of signals corresponds to the development of the Coronavirus disease (COVID-19) pandemic. On March 19, the State Department issued a Level 4 ``do not travel'' advisory, recommending that United States citizens avoid any global travel. Although this security measure emphasises international flights, it also influences domestic aerial connections. The continuous sequence of the signals in case of the MEWMA begins on March 21, 2020. In case of the MCUSUM, the start is on March 24. Although {\color{blue} in both cases} the control statistic resets to zero after each signalling, the repeated violation of the upper control limit is a clear indicator of this shift in network behaviour.

To identify smaller and more specific changes in the daily flight data of the US, one could also integrate nodal and edge covariates, which would refer to further aspects of the network. {\color{blue} Additionally}, control charts with smaller $k$ and $\lambda$ can be applied.

\section{Conclusion}

Statistical methods can be remarkably powerful for the surveillance of networks. However, due to the complex structure and potentially large size of the adjacency matrix, traditional tools for multivariate process control cannot directly be applied, as the network's complexity must be reduced first. For instance, this can be done by statistical modelling of the network. The choice of the model is crucial as it decides constraints and simplifications of the network which later influence the types of changes we are able to detect. In this paper, we show how multivariate control charts can be used to detect changes in {\color{blue} dynamic networks defined by a} TERGM. The proposed methods can be applied in real time. This general approach is applicable for various types of networks in terms of the edge direction and topology, as well as allows for the integration of nodal and edge covariates {\color{blue}and consideration of temporal dependence.}

The performance of our procedure is evaluated for different anomalous scenarios by comparing the CED of the calibrated control charts. According to the classification and explanation of anomalies provided by \cite{ranshous_shen_koutra_harenberg_faloutsos_samatova_2015}, the surveillance method presented in this work is applicable for event and change detection in temporal networks. 

Finally, we illustrated the applicability of our approach by monitoring daily flights in the United States. Both control charts were able to detect the beginning of the lock-down period due to the COVID-19 pandemic. The MEWMA chart signalled a change just two days after a Level 4 ``no travel'' warning was issued.

Despite the benefits of the TERGM, such as the incorporation of the temporal dimension and representation of the network in terms of its sufficient statistics, there are several considerable drawbacks. Other than the difficulty to determine a suitable combination of the network terms, the model is not suitable for networks of large size \citep{block2018change}. Furthermore, the temporal dependency statistics in the TERGM depend on the selected temporal lag and the size of the time window over which the data is modelled \citep{leifeld2015theoretical}. Thus, the accurate modelling of the network strongly relies on the analyst's knowledge about its nature. A helpful extension of the approach would be the implementation of the STERGM. In this case, it could be possible {\color{blue}to subdivide the network monitoring into two distinct streams}, so that the interpretation of changes in the network would become clearer.

{\color{blue} Another topic that demands additional research is the determination of cases when it is reliable to use the averaged network statistics $\hat{\bm{s}}_t$ to construct the monitoring procedure and not the parameter estimates $\hat{\bm{\theta}}_t$, as their estimation is more complex than of  $\hat{\bm{s}}_t$. Also, it would be beneficial to consider other estimators to compute $\hat{\bm{s}}_t$ and compare their effectiveness to detect anomalies.}

Regarding the multivariate control charts, there are also some aspects to consider. Referring to \cite{montgomery2012statistical}, the multivariate control charts perform well if the number of process variables is not too large, usually up to 10. Also, a possible extension of the procedure is to design a monitoring process when the values for $\bm{\Sigma}$ can vary between the in-control and out-of-control states. Whether this factor would beneficially enrich the surveillance remains open for future research. {\color{blue} Furthermore, we customise the application using} simulation methods to calibrate the charts. Hence, further development of adaptive control charts with different characteristics is interesting as they could remarkably improve the performance of the anomaly detection (cf. \citealp{sparks2019monitoring}).





\end{document}